\documentclass[]{spie}  

\usepackage{amsmath,amsfonts,amssymb}
\usepackage{graphicx}
\usepackage{caption}
\usepackage{subcaption}
\usepackage[colorlinks=true, allcolors=blue]{hyperref}
\usepackage{multirow}
\usepackage{amsmath,amsfonts,amssymb}
\usepackage{graphicx}
\usepackage[colorlinks=true, allcolors=blue]{hyperref}
\usepackage{todonotes}

\title{Systems design, assembly, integration and lab testing of WALOP-South Polarimeter}

\author[a*,b,c]{Siddharth Maharana}
\author[b,c,d*]{A. N. Ramaprakash}
\author[b]{Chaitanya Rajarshi}
\author[b]{Pravin Khodade}
\author[b]{Bhushan Joshi}
\author[b]{Pravin Chordia}
\author[b]{Abhay Kohok}
\author[b,k]{Ramya M. Anche}
\author[a]{Deepa Modi}
\author[c,e]{John A. Kypriotakis}
\author[b]{Amit Deokar}
\author[b]{Aditya Kinjawadekar}
\author[a,i]{Stephen B. Potter}
\author[c,e]{Dmitry Blinov}
\author[g]{Hans Kristian Eriksen}
\author[c,e]{Myrto Falalaki}
\author[a]{Hitesh Gajjar}
\author[h]{Tuhin Ghosh}
\author[g]{Eirik Gjerløw}
\author[c,e]{Sebastain Kiehlmann}
\author[c,e]{Ioannis Liodakis}
\author[c,e]{Nikolaos Mandarakas}
\author[f]{Georgia V. Panopoulou}
\author[c,e]{Vasiliki Pavlidou}
\author[d]{Timothy J. Pearson}
\author[c,e,j]{Vincent Pelgrims}
\author[c,l]{Anthony C. S. Readhead}
\author[l]{Raphael Skalidis}
\author[c,e]{Konstantinos Tassis}
\author[c,e]{Namita Uppal}
\author[g]{Ingunn K. Wehus}

\affil[a]{South African Astronomical Observatory, PO Box 9, Observatory, 7935, Cape Town, South Africa}
\affil[b]{Inter-University Centre for Astronomy and Astrophysics, Post bag 4, Ganeshkhind, Pune, 411007, India}
\affil[c]{Institute of Astrophysics, Foundation for Research and Technology-Hellas, Voutes, 70013 Heraklion, Greece}
\affil[d]{Cahill Center for Astronomy and Astrophysics, California Institute of Technology, Pasadena, CA, 91125, USA}
\affil[e]{Department of Physics, University of Crete, Voutes, 70013 Heraklion, Greece}
\affil[f]{Department of Space, Earth \& Environment, Chalmers University of Technology, 412 93 Gothenburg, Sweden}
\affil[g]{Institute of Theoretical Astrophysics, University of Oslo, P.O. Box 1029 Blindern, NO-0315 Oslo, Norway}
\affil[h]{School of Physical Sciences, National Institute of Science Education and Research, HBNI, Jatni 752050, Odisha, India}
\affil[i]{Department of Physics, University of Johannesburg, PO Box 524, Auckland Park 2006, South Africa}
\affil[j]{Université Libre de Bruxelles, Science Faculty CP230, Av. Franklin Roosevelt 50, 1050 Brussels, Belgium}
\affil[k]{Steward Observatory, University of Arizona, 933 N Cherry Ave, Tucson, Arizona, 85721, USA} 
\affil[l]{Owens Valley Radio Observatory, California Institute of Technology, MC 249-17, Pasadena, CA 91125, USA}

\authorinfo{Further author information: (Send correspondence to S.M. and A.N.R.)\\S.M.: E-mail: siddharth@saao.ac.za, A.N.R.: anr@iucaa.in}

\begin{document} 
\maketitle

\begin{abstract}
Wide-Area Linear Optical Polarimeter (WALOP)-South is the first wide-field and survey-capacity polarimeter in the optical wavelengths. On schedule for commissioning in 2024, it will be mounted on the 1 m SAAO telescope in Sutherland Observatory, South Africa to undertake the \textsc{PASIPHAE} sky survey. \textsc{PASIPHAE} program will create the first polarimetric sky map in the optical wavelengths, spanning more than 2000 square degrees of the southern Galactic region. In a single exposure, WALOP-South’s innovative design will enable it to measure the linear polarization (Stokes parameters $q$ and $u$) of all sources in a field of view (FoV) of $35\times35$ arc-minutes-squared in the SDSS-r broadband and narrowband filters between 500-750 nm with 0.1~\% polarization accuracy. 

The unique goals of the instrument place very stringent systems engineering goals, including on the performance of the optical, polarimetric, optomechanical, and electronic subsystems. In particular, the major technical hurdles for the project included the development of: (a) an optical design to achieve imaging quality PSFs across the FoV, (b) an optomechanical design to obtain high accuracy optical alignment in conjugation with minimal instrument flexure and stress birefringence on optics (which can lead to variable instrumental polarization), and (c) an on-sky calibration routine to remove the strong polarimetric cross-talk induced instrumental polarization to obtain 0.1\% across the FoV. All the subsystems have been designed carefully to meet the overall instrument performance goals.

As of May 2024, all the instrument optical and mechanical subsystems have been assembled and are currently getting tested and integrated. The complete testing and characterization of the instrument in the lab is expected to be completed by August 2024. While the instrument was initially scheduled for commissioning in 2022, it got delayed due to various technical challenges; WALOP-South is now on schedule for commissioning in second half of 2024. 

In this paper, we will present (a) the design and development of the entire instrument and its major subsystems, focusing the instrument's opto-mechanical design which has not been reported before, and (b) assembly and integration of the instrument in the lab and early results from lab characterization of the instrument’s optical performance.
\end{abstract}


\section{INTRODUCTION}\label{sec:intro}
Wide-Area Linear Optical Polarimeters (WALOPs) are two wide-field linear optical polarimeters currently being developed at the Inter-University Center for Astronomy and Astrophysics (IUCAA) in Pune, India. WALOP-South will be installed at the South African Astronomical Observatory's Sutherland Observatory on a 1 meter telescope\footnote{https://www.saao.ac.za/astronomers/1-0m/}, while WALOP-North will be mounted on a 1.3 meter telescope\footnote{http://skinakas.physics.uoc.gr/en/telescopes/tel\_130.html} at the Skinakas Observatory in Greece. These instruments will be used in the \textsc{PASIPHAE} survey\cite{tassis2018pasiphae}, which aims to cover 4,000 square degrees of the sky in the northern and southern Galactic polar regions and measure the polarization of approximately one million stars with a polarimetric accuracy better than 0.1\%. In contrast, current optical polarization catalogs include measurements for about 10,000 stars\cite{Heiles_2000}. The primary scientific goal of the \textsc{PASIPHAE} survey is to combine this highly accurate stellar polarimetric data with the GAIA survey's stellar distance measurements to create a 3D tomography map of the dust and magnetic field in the Milky Way Galaxy's polar regions\cite{Panopoulou_2019, Pelgrims2023}. For a detailed discussion of the scientific motivations and objectives, refer to the \textsc{PASIPHAE} program's white paper\cite{tassis2018pasiphae}.

\par The unique scientific objectives of the PASIPHAE survey drive the technical design goals for the WALOP instruments, applicable to both WALOP-North and WALOP-South, as detailed in Table~\ref{techtable}. The rationale and justification for the target values of each design parameter are explained in the optical design paper of the WALOP-South instrument by Maharana et al. 2021\cite{WALOP_South_Optical_Design_Paper}, hereafter referred to as Paper I.

\par Of the two WALOP instruments, WALOP-South is slated for commissioning first, in 2024. This manuscript details the overall design of the instrument, with a particular focus on the previously unreported optomechanical design. Sections~\ref{optical_design} and \ref{optomechanical_design_chapter} cover the optical and optomechanical designs of WALOP-South. The core instrument has been assembled in the lab at IUCAA, where it is undergoing testing and characterization for optical and polarimetric performance; we present early results from these efforts. Section~\ref{assembly} discusses the ongoing assembly, alignment, and testing efforts, while Section~\ref{conclusions} outlines the current status and next steps in the instrument's development.



\begin{table}[htbp!]
    \centering
    \caption{Design goals for WALOP-South instrument.}
    \label{techtable}
    \begin{tabular}{ccc}
        \hline
        \textbf{Sl. No}. & \textbf{Parameter} & \textbf{Technical Goal} \\
         \hline
        1 & Polarimetric Sensitivity & 0.05~\%\\
        2 & Polarimetric Accuracy & 0.1~\%\\
        3 & Polarimeter Type & Four Channel One-Shot Linear Polarimetry\\
        4 & Number of Cameras & 4 (One Camera for Each Arm)\\
        5 & Field of View & $30\times30$~arcminutes\\
        6 & Detector Size & $4k\times4k$ (Pixel Size = $15~{\mu}m$) \\
        7 & No. of Detectors & 4 \\
        8 & Primary Filter & SDSS-r \\
        9 & Imaging Performance & Close to seeing limited PSF \\
        10 & Stray and Ghost Light Level & Brightness less than sky brightness per pixel.\\
        \hline
    \end{tabular}
\end{table}

\section{Optical Design}\label{optical_design}

A major challenge in developing the WALOP-South instrument was creating an optical design that meets the requirements listed in Table~\ref{techtable}. Here, we present the overall optical model of the instrument and its predicted performance. While two/four-channel optical polarimeters with imaging on different detector areas have been created before, they have typically been designed for very narrow fields of view\cite{robopol,HOWPol} of around $1\times1$ arcminutes or have not been optimized for broadband filters\cite{salt_commisioning} (with widths greater than 100 nm). WALOP-South is the first of its kind: a wide-field, one-shot, four-channel imaging polarimeter with separate cameras for each channel. Its optical design has been tailored to work optimally with the 1 meter SAAO telescope's optics prescription and the specific temperature and observing conditions of the Sutherland Observatory, detailed in Table~\ref{telescope_details}.

\begin{table}[htbp!]
    \centering
    \caption{Telescope and site details of South African Astronomical Observatory's (SAAO) Sutherland Observatory.}
    \label{telescope_details}
    \begin{tabular}{cc}
    \hline
    \textbf{Parameter} & \textbf{Value} \\
    \hline
    Telescope Type & Cassegrain Focus and Equatorial Mount \\
    Primary Mirror Diameter & 1~m \\
    Secondary Mirror Diameter & 0.33~m \\
    Nominal Telescope f-Number & 16.0 \\
    Altitude & 1800~m\\
    Median Seeing FWHM & 1.5" \\
    Extreme Site Temperatures & $-10^{\circ}~C$ to $40^{\circ}~C$\\
    \hline
    \end{tabular}
\end{table}

\begin{figure}
    \centering
\begin{subfigure}{0.49\textwidth}
    \centering
    \fbox{\includegraphics[scale=0.235]{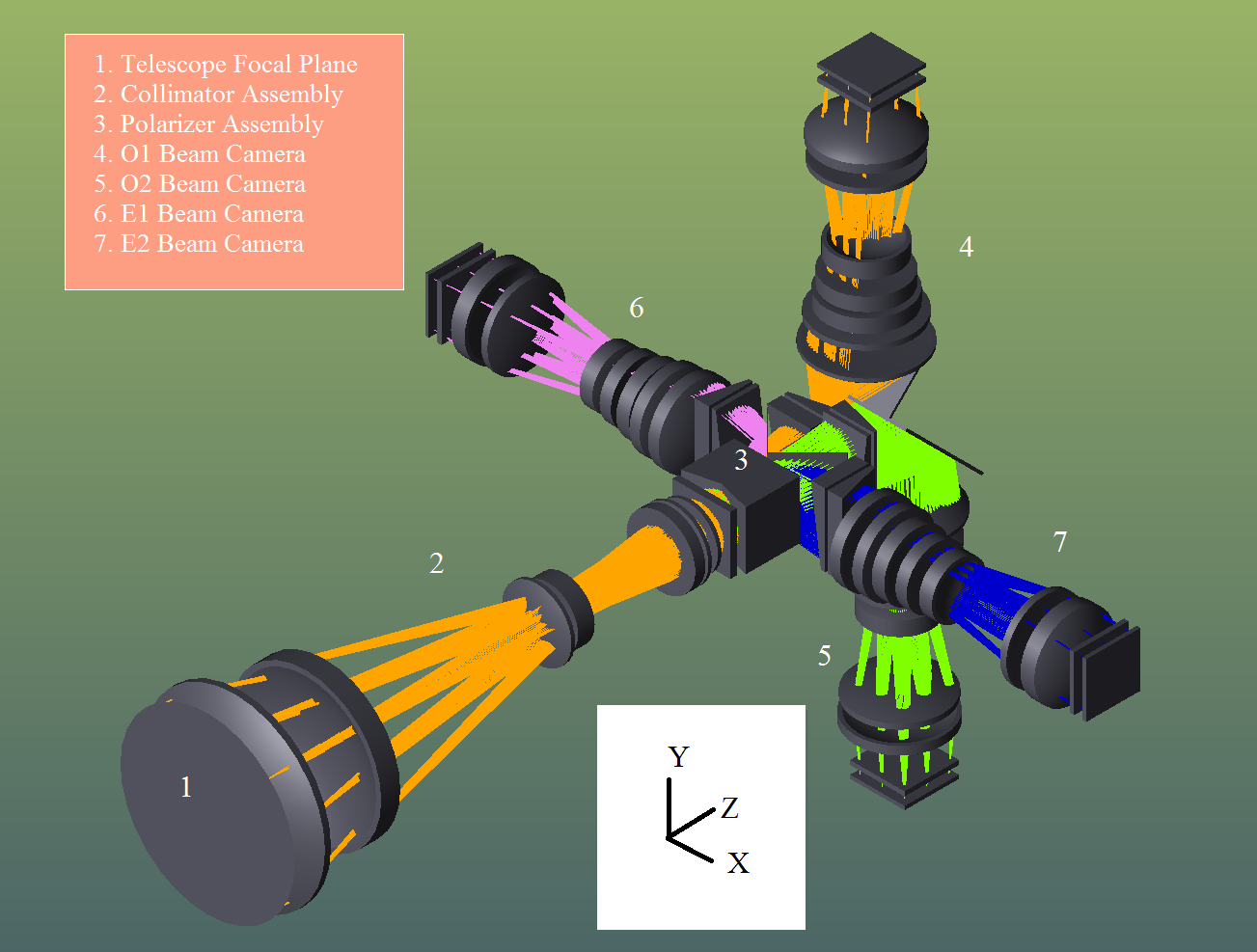}}
    \caption{The optical model of the WALOP-South instrument features a polarizer assembly that splits the pupil beam into four channels and directs them along the ±x and ±y directions. Each channel is equipped with its own camera to capture the entire field of view on a 4k × 4k CCD detector.}
    \label{WALOP-S}
\end{subfigure}
\begin{subfigure}{0.49\textwidth}
    \centering
    \frame{\includegraphics[scale = 0.185]{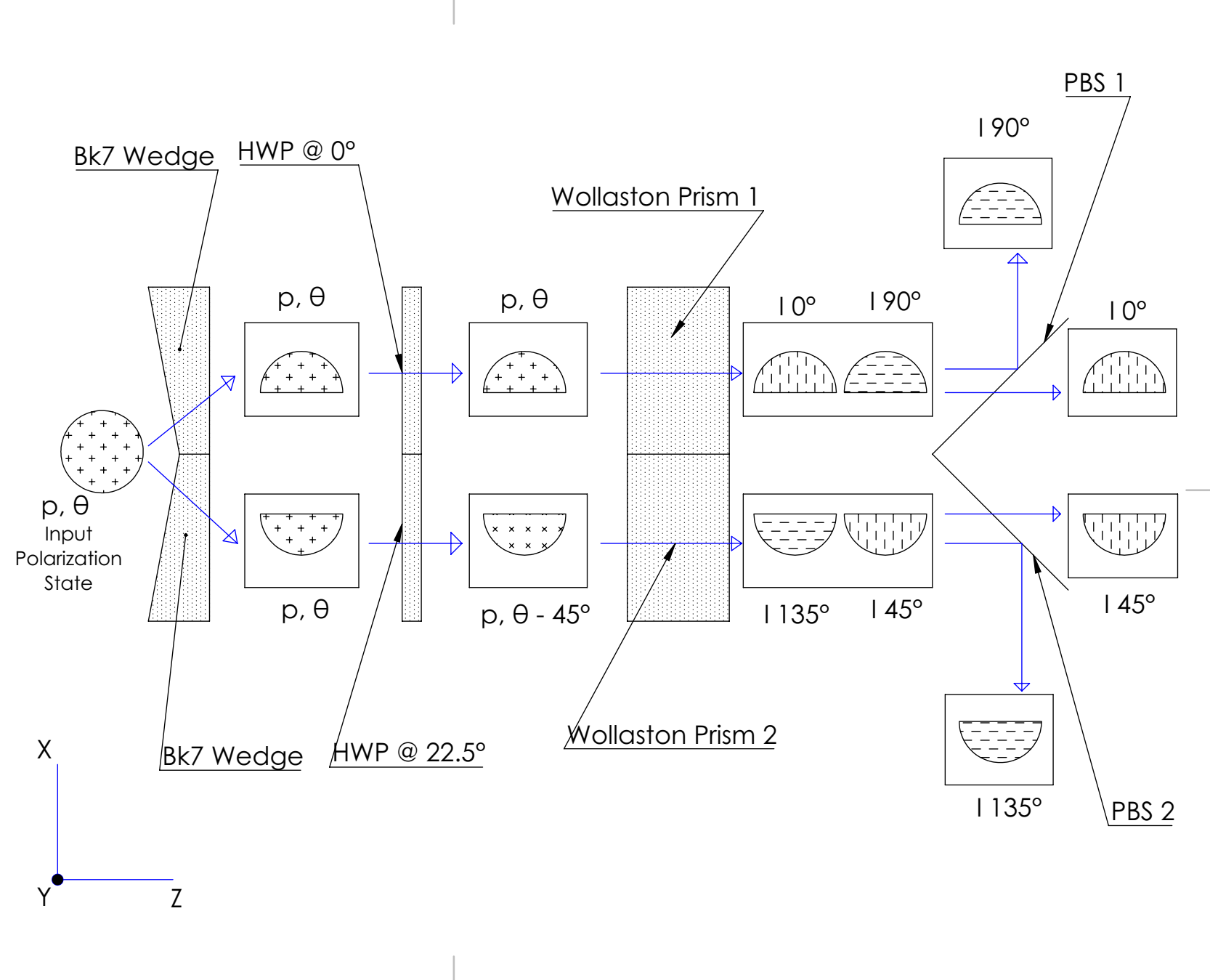}}
    \caption{A cartoon drawing depicting the polarizer assembly of the WALOP-South instrument, showing $p$ and $\theta$ as viewed from the x-y plane when looking along the z-axis. The modulation in the polarization of the beams as they go through this system is annotated.}
    \label{pol_ass_cartoon}
\end{subfigure}
    \caption{Details of the optical model of WALOP-South and the working of the polarizer assembly.}
    \label{Details of the optical model of WALOP-South}
\end{figure}

The optical model of WALOP-South was designed and analyzed using the Zemax optics software. Figure~\ref{WALOP-S} illustrates the optical model of the instrument. The entire optical system consists of the following assemblies: a collimator, a polarizer assembly, and four cameras (one for each channel). The collimator assembly begins at the telescope focal plane, aligned along the z-axis, creating a pupil image that is fed into the polarizer assembly.

The polarizer assembly functions as the polarization analyzer system of the instrument, splitting the pupil beam into four channels corresponding to $0^{\circ}$ , $45^{\circ}$, $90^{\circ}$, and $135^{\circ}$ polarization angles, referred to as O1, O2, E1, and E2 beams, respectively. Additionally, this assembly folds and steers the O beams along the +y and -y directions and the E beams along the +x and -x directions. Each channel has its own camera to image the entire FoV on a 4k × 4k CCD detector. Although the required field of view was 30 × 30 arcminutes, the obtained field of view of the instrument is 34.8 × 34.8~arcminutes$^{2}$. Table~\ref{op_design_summary} lists the design parameters of the optical system.

The polarizer assembly is the most novel and complex aspect of WALOP-South's optical design, and its architecture and functionality is shown in Figure~\ref{pol_ass_cartoon} and Section~\ref{pol_ass}, with more details presented in Paper I. As part of the optical design, we also developed a guider camera (Figure~\ref{collimator_camera_barrel_cross_section}) for the instrument and new baffles for the telescope to accommodate the large FoV of WALOP-South. These additions are also described in Paper I.

WALOP-South uses two large aperture (80 × 45 mm) calcite Wollaston Prisms (WP) as polarization analyzers. WPs are employed in WALOPs, as in most polarimeters, due to their high extinction ratios ($> 10^{5}$) and nearly symmetrical angular splitting of orthogonal polarization states. While double WPs have been used in astronomical polarimeters before, such as in HOWPol and RoboPol, these instruments have smaller FoV for which WPs with split angles of around $1^{\circ}$ are sufficient. However, for WALOPs' field size, WPs with large split angles in the range of $5^{\circ}$ to $10^{\circ}$ are necessary.

These large split-angle WPs introduce significant spectral dispersion in the split beams when using broadband filters due to the wavelength dependence of the split angle. This is in addition to the typical problems of large aberrations of off-axis objects caused by the very wide field. The optical design of WALOP-South overcomes these challenges and achieves a point spread function (PSF) close to the seeing limit across the FoV.

\subsection{Polarizer Assembly Design}\label{pol_ass} 

The system consists of four sub-assemblies: (a) Wollaston Prism Assembly (WPA), (b) Wire-Grid Polarization Beam-Splitter (PBS), (c) Dispersion Corrector Prisms (DC Prisms), and (d) Fold Mirrors.

The WPA comprises two identical calcite WPs, each paired with a half-wave plate (HWP) and a BK7 glass wedge positioned in front of it (Figure~\ref{pol_ass_cartoon}). These WPs have an aperture of 45 × 80 mm and a wedge angle of $30^{\circ}$, resulting in a split angle of $11.4^{\circ}$ at $0.6~{\mu}m$ wavelength. The left WP is equipped with an HWP whose fast axis is aligned at $0^{\circ}$ relative to the instrument coordinate system, to separate $0^{\circ}$ and $90^{\circ}$ polarizations, while the right WP has an HWP with a fast axis at $22.5^{\circ}$ to split the $45^{\circ}$ and $135^{\circ}$ polarizations. The BK7 wedges at the beginning of the WPA equally share the incoming pupil beam and ensure that rays from off-axis objects in the FoV, entering at oblique angles, do not hit the interface between the WPs. This prevents throughput loss and instrumental polarization from scattering at the surface. Thus, the WPA separates the input beam into O1, O2, E1, and E2 beams, corresponding to polarization angles of $0^{\circ}$, $45^{\circ}$, $90^{\circ}$ and $135^{\circ}$, respectively.

The PBS acts as a beam selector, allowing both O beams to pass through while folding the E1 and E2 beams along the -x and +x directions. Figure~\ref{pol_ass_cartoon} illustrates the overall functioning of the WPA and PBS components of the polarizer assembly. The DC Prisms, a pair of glass prisms located in the path of each of the four beams after the PBS, correct the spectral dispersion introduced by the WPA. Additionally, mirrors placed at $\pm~45^{\circ}$ to the y-z plane fold the O beams into +y and -y directions, limiting the instrument's length to 1.1 m from the telescope focal plane.

\begin{table}[htbp!]
    \centering
    \caption{Key parameters of the WALOP-South Optical Design.}
    \label{op_design_summary}
    \begin{tabular}{cc}
        \hline
        \textbf{Parameter} & \textbf{Design Value/Choice}\\
        \hline
        Filter & SDSS-r \\
        Telescope F-number & 16.0 \\
        Camera F-number & 6.1 \\
        Collimator Length & 700~mm\\
        Camera Length & 340~mm \\
        No of lenses in Collimator & 6 \\
        No of lenses in Each Camera & 7 \\
        Detector Size & $4096\times4096$ \\
        Pixel Size & $15~{\mu}m$ \\
        Sky Sampling at detector & 0.5"/pixel\\
        \hline
        
    \end{tabular}

\end{table}

The entire WPA is cemented together as a single unit using Norland 65 cement, a flexible adhesive chosen specifically to withstand the significant stresses in the WPs due to the anisotropic thermal expansion of calcite WPs\cite{Pellicori:70} under the large temperature variations expected at SAAO’s Sutherland Observatory. To ensure the cement's flexibility, sacrificial calcite WPs bonded with Norland 65 and other candidate cements were tested in an environmental chamber at temperatures ranging from -10 to 40 degrees Celsius. Details of these tests on the candidate cements can be found in Appendix~\ref{cement_tests}.

\subsection{Optical Performance and Tolerance Analysis}

The spot diagrams and predicted PSFs are very similar for all four beams. The O1 and O2 beams share the same optical paths, resulting in identical spot diagrams. The same applies to both E1 and E2 beams. The averaged RMS (root mean squared) radii for the O and E beams, shown in Table~\ref{MC_sim}, are 11.63 and 11.77 ${\mu}m$, respectively. In comparison, for a Gaussian beam with a FWHM of 1.5 arcseconds (median seeing value at Sutherland Observatory), the RMS radius  at the detector measures 19.1~${\mu}m$.

Using Zemax software, a thorough tolerance analysis on the optical system was performed utilizing Monte Carlo simulations to predict the potential degradation in spot sizes for the instrument, as well as determining the necessary tolerances for manufacturing the optical and mechanical components of the system. Two compensators were identified: (a) the separation between the primary and secondary telescope mirror, and (b) the distance between the final camera lens and each camera's detector. The outcomes of 20,000 MC simulations are presented in Table~\ref{MC_sim}. The average RMS spot size for the O and E beams derived from the simulations are 17.1 and 15.37~${\mu}m$ respectively each, smaller compared to the RMS radius of a 1.5 arcsecond FWHM Gaussian beam at the sensors (19.1~${\mu}m$). So we anticipate achieving a nearly seeing-limited PSF on the detectors (for a detailed and quantified prediction of the instrument's anticipated imaging capabilities, see Paper I). The tolerances needed for aligning the optical assembly, as shown in Table~\ref{mech_toleraces}, were used to generate the MC simulation results. The tolerance values are identical for each respective optical lens/prism in all four beams.


\subsection{Polarimetric Modelling}

\par Wide-field polarimeters like WALOP-South are prone to high instrumental polarization values, and consequently, the measured Stokes parameters may be significantly different from the real values. As the design goal for the instrument is to achieve polarimetric measurement accuracy of 0.1~\% over the entire FoV, these need to be modeled and then corrected from the measurements. In Maharana et al. 2022\cite{WALOP_Calibration_paper}, referred to as Paper II from hereon, the complete polarimetric modeling of WALOP-South is presented, characterizing the amount and sources of instrumental polarization. For this work, detailed and very accurate polarimetric modeling analysis software was custom developed by the instrument team that can be used with optical design software like Zemax for the characterization of WALOP-like instruments. We find that the main source of instrumental polarization in WALOPs arises from the angle of incidence-dependent retardance by the HWPs in the WPA.

To estimate and correct for these effects in order to accurately recover the real Stokes parameters, we have developed a detailed method for the calibration of the instruments. The standard procedure involves creating a mapping function between the instrument-measured and real Stokes parameters of a source, typically in the form of a matrix. This requires known polarized sources to determine the coefficients of mapping functions. Because there are not many on-sky wide-field calibration sources available, a calibration linear polarizer (CLP) is included in the instrument before any polarization-inducing optics, such as lenses. The CLP is placed on a motorized rotary stage: for calibration observations it is inserted in the optical path using a linear mechanism and taken out for the main science observations. When observing any stellar field, the CLP will supply the instrument with fully linearly polarized light at different Electric Vector Position Angles (EVPA). Using this calibration model and simulated data, we show how to correact the instrumental polarization to achieve 0.1\% accuracy of $p$ across the entire FoV. Through the utilization of the CLP, polarimetric wide-field flats such as sky adjacent to the bright Moon\cite{sky_polarization} (described below), and standard stars\cite{RoboPol_standards}, we can calibrate the FoV of WALOP-South to within 0.1\% accuracy. Furthermore, we have verified and proven the effectiveness of the calibration technique by applying it to a table-top WALOP-like test-bed polarimeter in the lab.
\par As part of developing the calibration method for WALOP-South, we have shown that the sky around the Full Moon within a range of two days can serve as a highly effective polarimetric flat source (with an accuracy greater than 0.05\% in $p$) for wide-field polarimeter calibration. Observations conducted with the RoboPol instrument have confirmed this; in Maharana et al. \cite{sky_polarization}, we present these observations and results. In response to the need for a large pool of reliable polarimetric standard stars for the calibration of WALOP polarimeters and lack of trustworthy polarized and unpolarized calibration standard stars in optical polarimetry community, a new catalog of 65 stable (at levels of 0.1~\% or better) standard stars has been created after a 5-year monitoring campaign of 107 stars using the RoboPol polarimeter\cite{RoboPol_standards}.

\begin{table}[ht!]
\centering
\caption{Findings from tolerance analysis using Monte Carlo (MC) simulations. The Root-Sum-Square (RMS) spot radius is calculated by adding the offset caused in nominal spot radius by all the mechanical and optical tolerances in quadrature.}
\label{MC_sim}
\begin{tabular}{ccc}
\hline
Parameter                & O-Beams             & E-Beams             \\ 
                         & RMS Spot Radius (${\mu}m$) & RMS Spot Radius (${\mu}m$) \\ \hline
Nominal Spot             & 11.63               & 11.77               \\ 
Root-Sum-Square (RMS)      & 17.1                & 15.37               \\ 
MC Simulation Best Case  & 11.72               & 11.7                \\ 
MC Simulation Worst Case & 37.4                & 25.5                \\ 
MC Simulation Mean       & 17.54               & 15.72               \\ 
MC Simulation Std Dev    & 0.003               & 0.0018              \\ \hline
\end{tabular}
\end{table}

\begin{table}[htbp!]
\centering
\caption{Alignment tolerances for the optical components of the WALOP-South instrument. The values are common for corresponding elements in all the four cameras.}
\label{mech_toleraces}
\begin{tabular}{cccc}
\hline
\multirow{1}{*}{Lens   Name} & \multirow{1}{*}{Decentre~(${\mu}m$)} & \multirow{1}{*}{Axial~(${\mu}m$)} & \multirow{1}{*}{Tilt~(arcminute)} \\ \hline
Collimator Lens 1            & 50                                       & 200                                   & 3                                        \\ 
Collimator Lens 2            & 50                                       & 200                                   & 3                                        \\ 
Collimator Lens 3            & 50                                       & 200                                   & 3                                        \\ 
Collimator Lens 4            & 30                                       & 200                                   & 2                                        \\ 
Collimator Lens 5            & 50                                       & 200                                   & 1                                        \\ 
Collimator Lens 6            & 20                                       & 100                                   & 2                                        \\ 
Camera Lens 1                & 30                                       & 50                                    & 1                                        \\ 
Camera Lens 2                & 30                                       & 30                                    & 1                                        \\ 
Camera Lens 3                & 30                                       & 50                                    & 1                                        \\ 
Camera Lens 4                & 30                                       & 200                                   & 1                                        \\ 
Camera Lens 5                & 30                                       & 200                                   & 2                                        \\ 
Camera Lens 6                & 50                                       & 200                                   & 3                                        \\ 
Camera Lens 7                & 50                                       & 200                               & 3                                        \\ 
WPA                          & 50                                       & 100                                   & 5                                        \\ 
PBS                          & 50                                       & 100                                   & 5                                        \\ 
DC Prism 1                   & 50                                       & 100                                   & 5                                        \\ 
DC Prism 2                   & 50                                       & 100                                   & 5                                        \\ 
Fold Mirror                  & 50                                       & 100                                   & 5                                        \\ \hline
\end{tabular}

\end{table}

\section{Optomechanical Design of WALOP-South}\label{optomechanical_design_chapter}

\par The optical model of the WALOP-South instrument is unique for polarimeters, owing to its four cameras, sophisticated polarizer assembly, and the large number of optical elements (in total, there are 50 lenses, prisms and mirrors). Being a wide-field camera system, WALOP-South instrument needed to be designed to meet stringent optical performance requirement of near seeing limited PSF over the full FoV, leading to tight optical alignment requirements. The instrument has multiple motion mechanisms, as well a custom designed guider camera. The optomechanical model for the instrument was developed taking into account all these considerations. The technical design requirements for the optomechanical model of the WALOP-South instrument are:





\begin{enumerate}
    \item To hold, align and assemble all the optical elements to accuracies listed in Table~\ref{mech_toleraces}. In particular, the lenses at either side of the pupil in the collimator and the camera assemblies have the most stringent alignment accuracy requirements: tilt and decenter of 20-30~${\mu}m$. 
    
    \item Maintain the alignment of the optics for the all possible pointings/inclinations of the telescope from zenith to up to $30^{\circ}$ from the horizon, i.e. airmass of 2, as most observations will be done in this telescope pointing window. Long and bulky instruments are susceptible to misalignment due to flexure coming from the telescope and the instrument. For example, during the commissioning of the WIRC+Pol polarimeter \cite{Tinyanont2018}, the instrumental polarization was found to be varying with telescope inclination.
    
    \item Lens holders should exert minimal stress ($< 7~\rm{MPa}$) on the glasses due to mounting method and temperature variations at the observatory. Stress in glass causes stress birefringence and resulting retardance, leading to instrumental polarization in the light passing through the glass.
    
    \item Special care has to be taken so that no stray light reaches the detectors, especially light of any one of the polarization type (O1, O2, E1 and E2 beam) should not reach a detector other than its own. Performance of polarimeters, especially with FoV are very susceptible to stray and ghost light contamination, and can lead to spurious polarization signals. 
    
    \item The maximum mass of instrument that can be mounted on the SAAO 1~m telescope is 200 Kgs; so the instrument mass should be under 180 Kg, with a contingency of 20 Kgs. 
    
    \item The instrument requires controlled motions for numerous subsystems; Table~\ref{control_goals} contains all motion systems, while Section~\ref{mechanisms} provides the working explanation for each. The optomechanical system must have the necessary mechanisms to ensure all movements are accurate as needed.
    
    \item The instrument design needs to include provisions for installing all necessary electrical connectors and control boxes, including detector and motion control boxes.
\end{enumerate}

\subsection{Overall Optomechanical Model}
\par
\begin{figure}
    \centering

\begin{subfigure}{0.99\textwidth}
\centering
\includegraphics[scale=0.5]{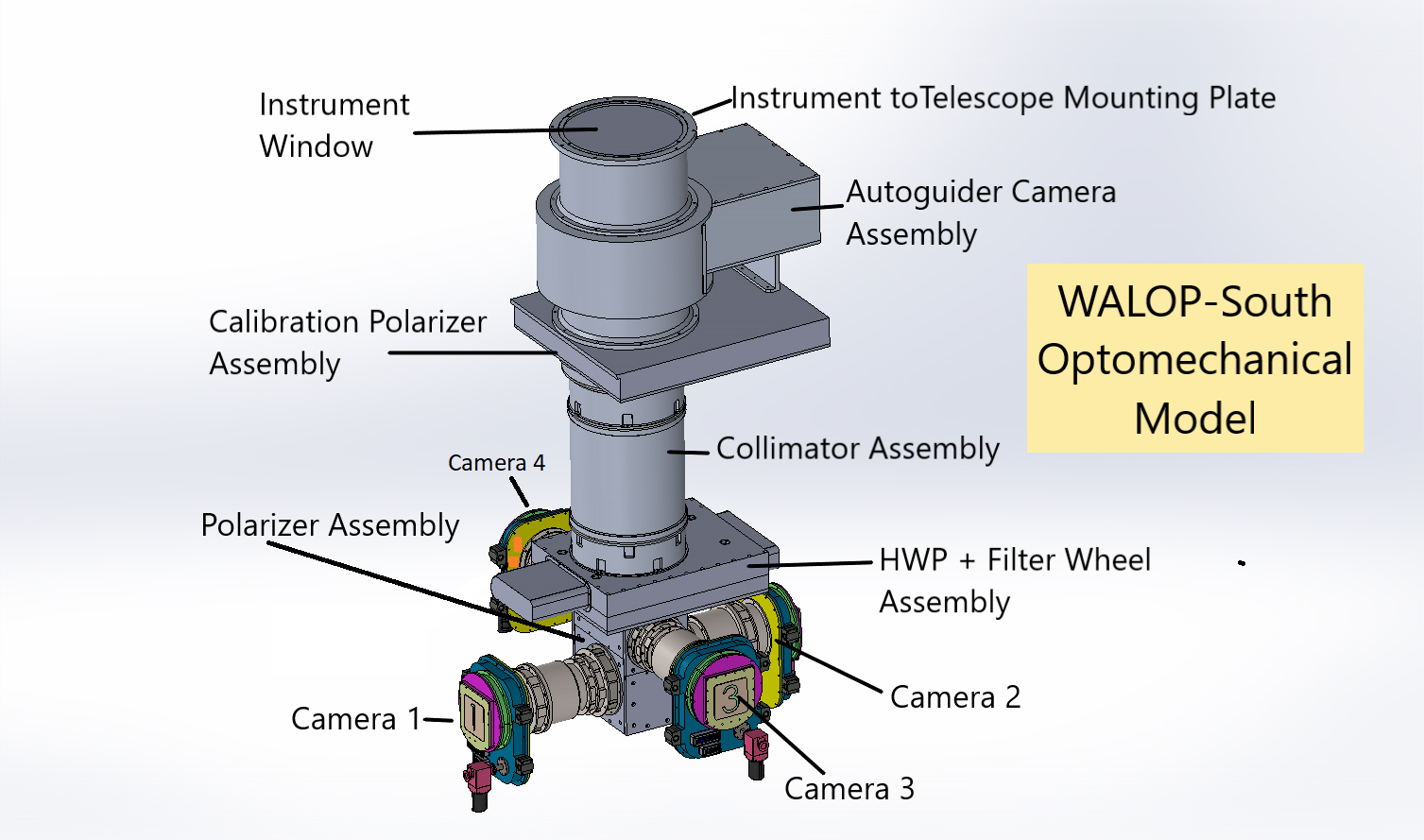}
\caption{The overall optomechanical model of the WALOP-South instrument, without electrical connectors and control boxes mounted. The various major subsystems in the model have been marked.}
\label{optmech_shaded}
\end{subfigure}

\begin{subfigure}{0.99\textwidth}
\centering
\includegraphics[scale=0.45]{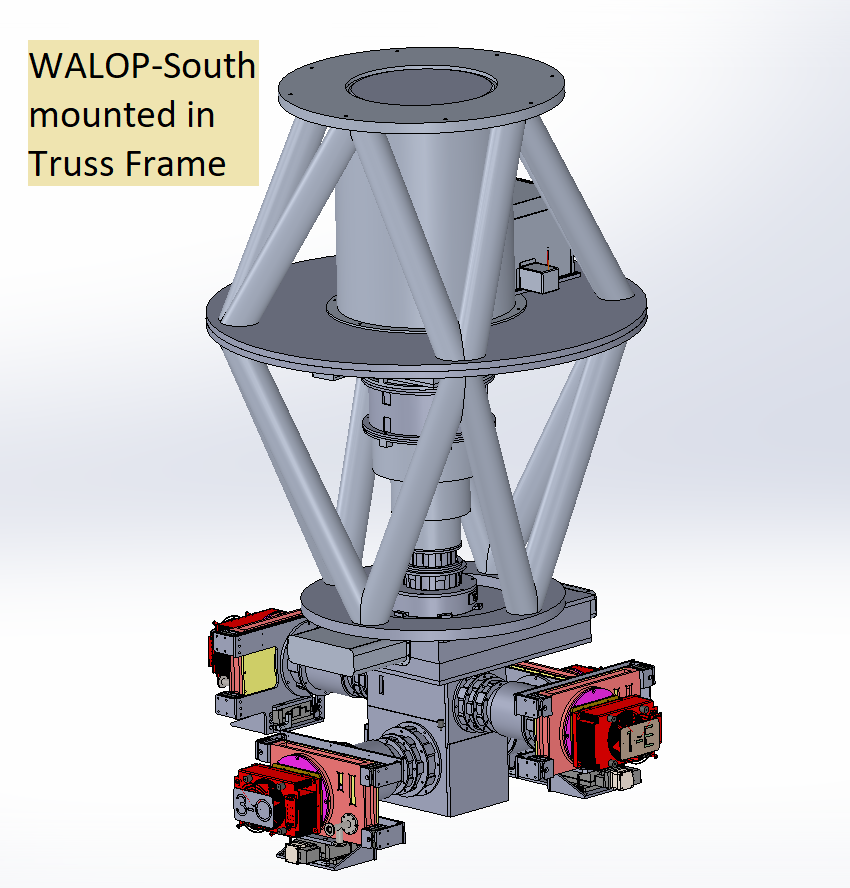}
\caption{The overall optomechanical model of the WALOP-South instrument mounted inside its truss structure.}
\label{optmech_shaded_with_truss}
\end{subfigure}

\caption{Complete optomechanical model of the WALOP-South instrument.}
\label{overall_optomech_model}
\end{figure}

 The SAAO 1~m telescope is a Cassegrain type mirror system on a German equatorial mount. The instrument model has been designed to operate optimally at the telescope and site conditions of the Sutherland Observatory (Table~\ref{telescope_details}). The optomechanical design and analysis work was done in SolidWorks software. Figure~\ref{optmech_shaded} depicts the complete optomechanical model of the device, excluding trusses, electrical connections, and mounted control boxes. The various parts of the instrument are labeled in the picture. Figure~\ref{optmech_shaded_with_truss} shows the instrument model mounted on the trusses designed to control the instrument flexure (described later in this section).

\par The instrument begins with an instrument window located immediately following the mounting flange to create a physical barrier from the external surroundings, keeping dust out and preventing it from settling on optical surfaces, which could result in false polarization signals. Before the main instrument optics (i.e., the collimator), there are the auto-guider camera and CLP sub-assemblies. The primary optomechanical structure of the device can be categorized into these subsystems- (a) collimator, (b) polarizer and (c) four camera assemblies. The collimator and camera assemblies are shaped like barrels (Section~\ref{barrel_design}), while the polarizer assembly, which includes the Wollaston Prism Assembly, wire-grid Polarization Beam-Splitters, Dispersion Corrector Prisms, and the Fold Mirrors, are contained within a box from which the four camera barrels extend in different directions. Additionally, between the collimator and the polarizer assembly, near the pupil, lies two mechanisms- the filter mechanism as well as the calibration half-wave plate (HWP) mechanism. 

\par As already noted, special care has to be taken to ensure that no stray light reaches the detectors. For this, all lens mounts and spacers have been black anodized. At multiple places throughout the instrument, apertures of the size of the beam footprint area have been made to block stray light rays outside this footprint area. 

\subsection{Collimator and Camera Barrel Design}\label{barrel_design}

\begin{figure}[htbp!]
\centering
\begin{subfigure}{0.49\textwidth}
    \centering
    \fbox{\includegraphics[scale = 0.225]{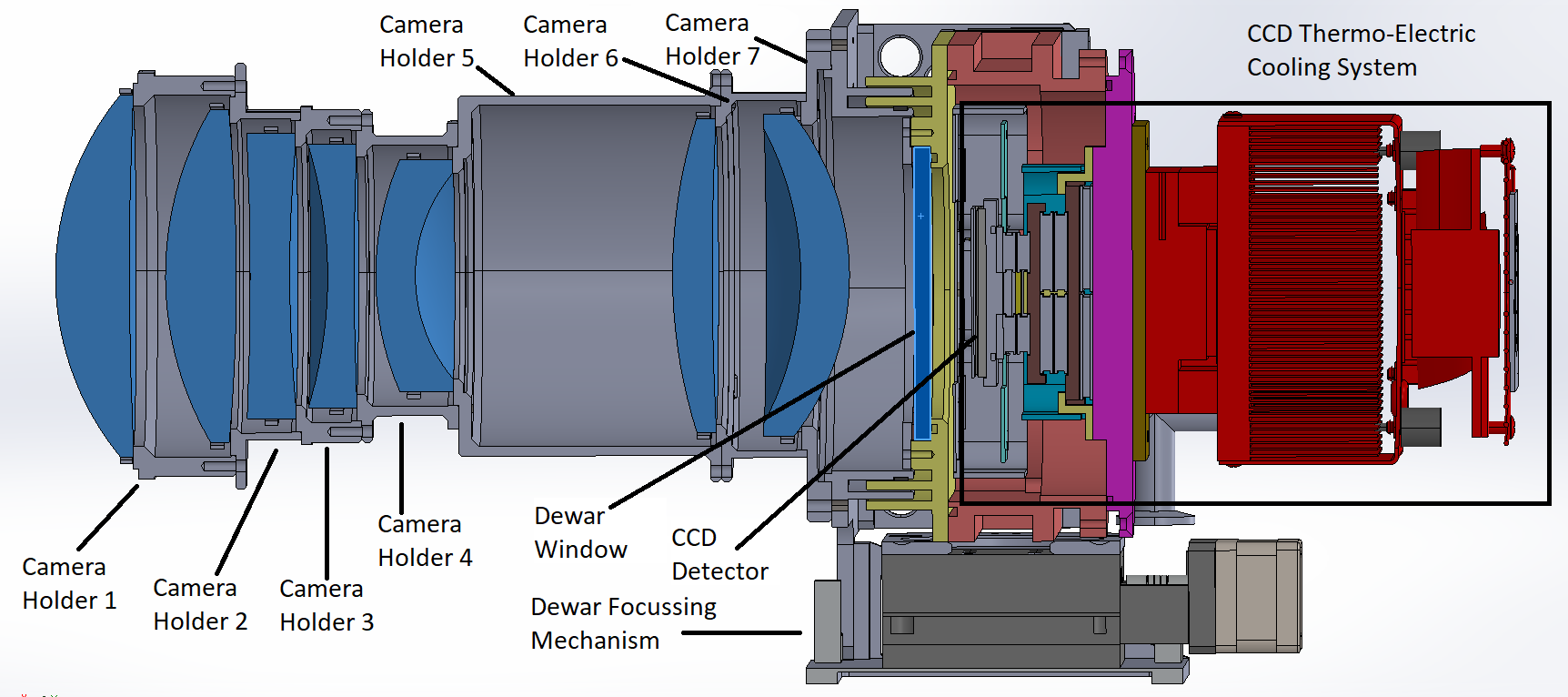}}
\end{subfigure}
\begin{subfigure}{0.49\textwidth}
    \centering
    \fbox{\includegraphics[scale = 0.225, angle = 0]{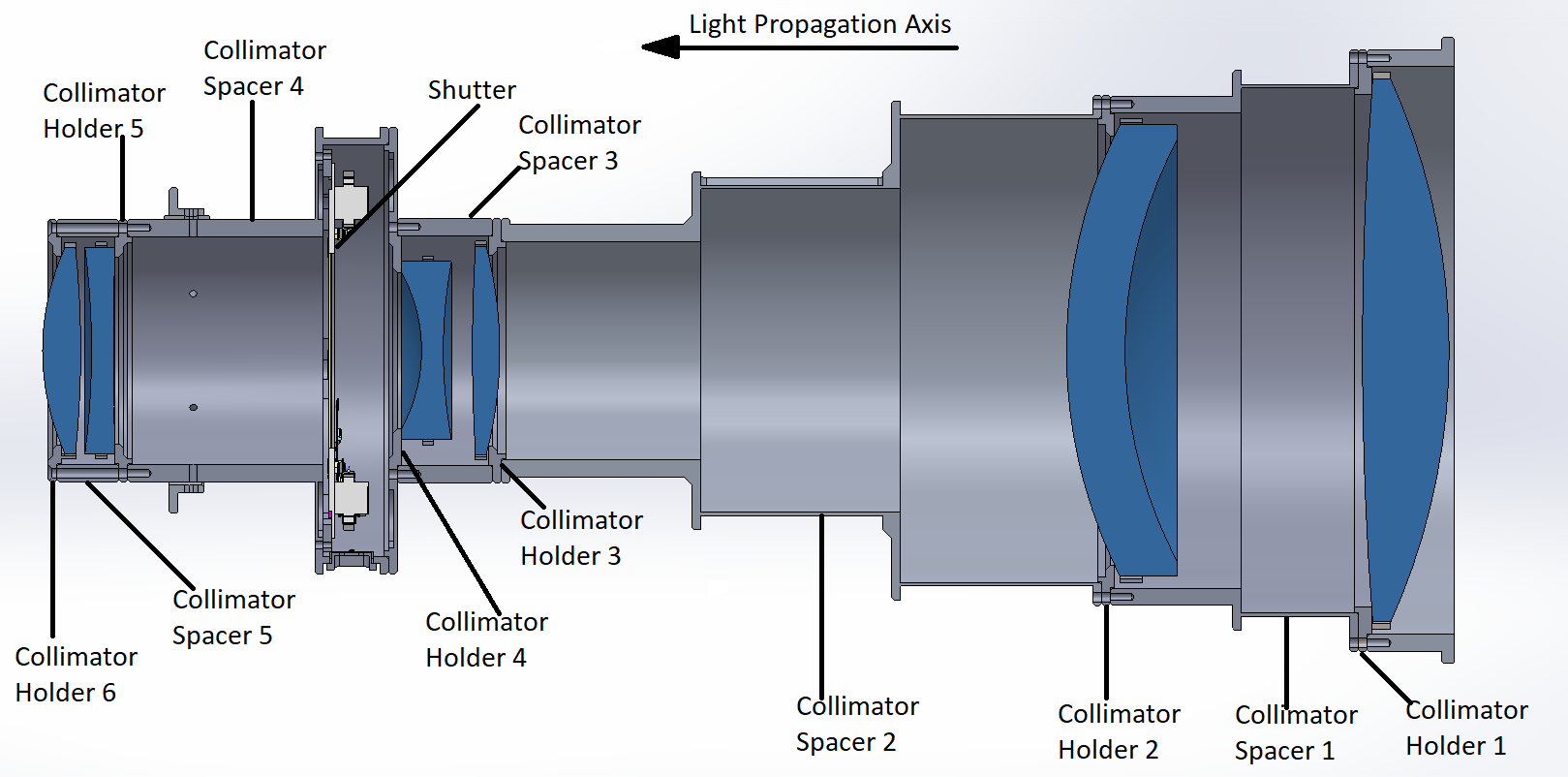}}
\end{subfigure}
\caption{Left:Cross-section of one of the four camera barrels. Every lens is held in it's holder, which is then connected to the succeeding lens holder. In this barrel, to obtain better alignment accuracy, the lens holder and spacer for all the individual lenses have been integrated into one component. Right:Cross-section of the collimator assembly barrel. Every lens is held in it's holder, which is then connected to the succeeding lens holder through a spacer element. The common shutter for all the four cameras also lies inside this assembly. }
\label{collimator_camera_barrel_cross_section}
\end{figure}

Lens assemblies are often designed in form of barrels as they provide high stiffness along with accurate alignment obtained through axial mating of successive lens holders. We have used the same approach for WALOP-South instrument. Figure~\ref{collimator_camera_barrel_cross_section} shows the cross section of one of the camera barrels as an example. L-shaped steps are accurately made in each lens holder with high precision machining (machining tolerance of 20-25 ${\mu}m$) on which the inner diameter of the the next lens holder sits. This interface is then tightened by screws coming from one holder and mating with corresponding thread on the other holder. For example, in Figure~\ref{collimator_camera_barrel_cross_section}, M4 bolts will sit on the Camera Lens 2 holder, pass through a clear hole in it and get connected to the M4 thread holes in Camera Lens 1 holder. The same approach has been used in all the other assemblies. Figure~\ref{collimator_camera_barrel_cross_section} also shows the cross-section of the collimator assembly. Due to large air separations between successive lenses in the collimator, cylindrical barrels/spacers are used to connect the lens holders.

\par
\begin{figure}
\begin{subfigure}{0.49\textwidth}
    \centering
    \frame{\includegraphics[scale = 0.1675]{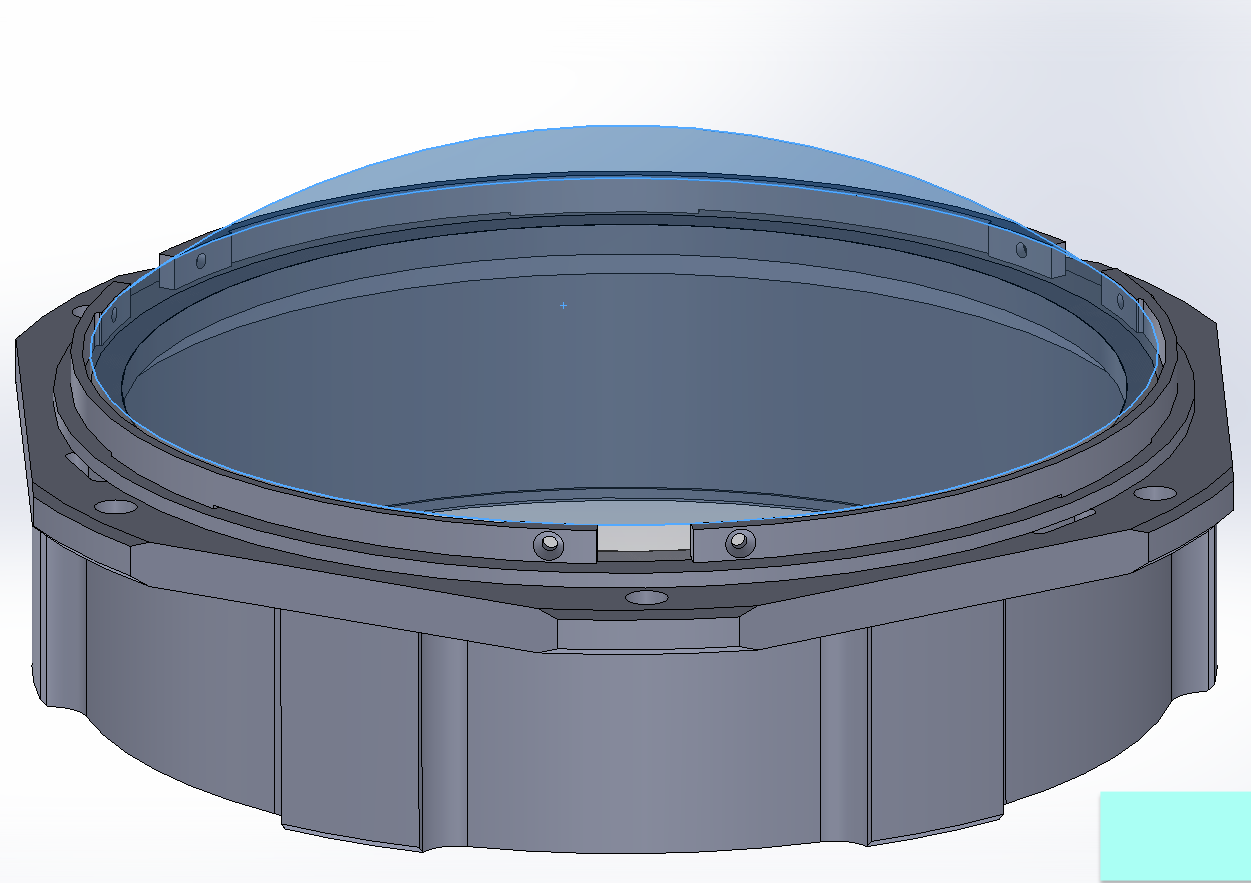}}
    \caption{CAD model based lens mounted on its lens holder.}
    \label{CameraL1_t}
\end{subfigure}
\begin{subfigure}{0.49\textwidth}
    \centering
    \includegraphics[scale = 0.425]{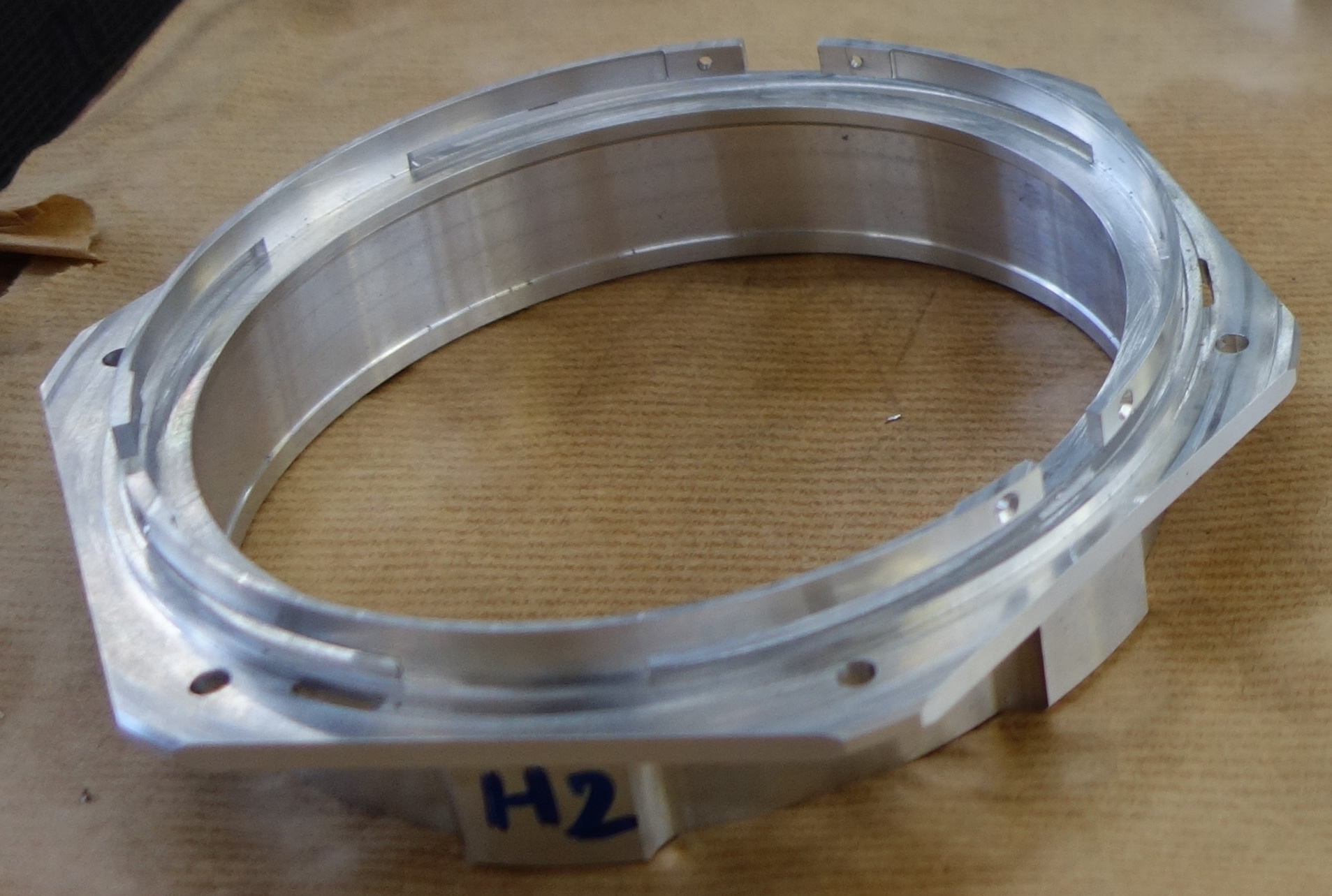}
    \caption{Fabricated lens mount (before anodization).}
\end{subfigure}
\caption{Design of lens mount for smaller aperture lenses ($\phi<$ 200~mm), used for all lenses except collimator lenses 1 and 2. It consists of 6 flexure arms around the lens' cylindrical rim..}
\label{CameraL1_mount}
\end{figure}

 Flexure based lens mounts are commonly utilized for ensuring precise alignment of the lenses within their respective holders\cite{Vukobratovich_flexure, Yoder1}.  Figure~\ref{CameraL1_mount} shows one of the lens holder mounts of the instrument. The design of these mounts were obtained from Finite Element Analysis (FEA) based simulations in the SolidWorks software. The lens is attached to the holder using the flexure arms, whose tips are bonded with the rim of the lens at multiple points using an adhesive. This constrains all the degrees of freedom of the lens, expect radial expansion and contraction in the plane of mounting, thus maintaining the alignment irrespective of relative contraction/expansion between the lens and and lens holder due to temperature changes. Additionally, these mounts are designed to maintain the centering of the lens to within 4-5 microns with respect to its holder at all orientations, especially when the lens mount is pointing towards the horizon (the worst case scenario), as will be the nominal orientation of the camera lenses when the instrument is pointing to the zenith. 

The lens mount material is Aluminium-6061 alloy for all lenses, except for the largest lenses (collimator lens 1 and 2), which use Titanium 6Al-4V alloy. The thermal expansion coefficient of Titanium 6Al-4V alloy is $8.9\times10^{-6}/K$, similar to most glasses ($7-10\times10^{-6}$), whereas the thermal expansion coefficient of Aluminium-6061 alloy is $24\times10^{-6}$, much higher than most glasses. Despite being suitable for lens mounts to alleviate stress caused by temperature fluctuations, Titanium 6Al-4V is 60~\% heavier (with a density of $4.4\times10^{3}~{Kg/m^3}$) compared to Aluminium-6061 alloy ($2.7\times10^{3}~{Kg/m^3}$). Furthermore, Aluminium-6061 is cheaper and readily available, and, most significantly, simpler to manufacture to the precise tolerances needed in our case. Therefore, except for the first two collimator lenses which are anticipated to experience higher thermal stresses because of their larger apertures ($>~20~\rm{cm}$), all other lens mounts are constructed from Aluminium-6061 alloy. In order to decrease the strain on the two lenses, 8 flexure arms are added, while 6 flexure arms are enough for the rest of the lenses. Figure~\ref{CameraL1_mount} also shows the actual lens mount (before black anodization) fabricated at the SAAO workshop. The decision on the material and number of flexure arms for each lens mount was taken based on mechanical stress and consequent birefringence studies on the optical lenses presented in Anche et al.\cite{walop_stress_birefringence}.



\subsection{Polarizer Assembly Design}
The polarizer assembly begins near the re-imaged pupil of the telescope after the collimator. All the optical components of the assembly have no optical power being flat surfaces and have collimated beam of light as input and output. Consequently, these do not have stringent alignment requirements (Table~\ref{mech_toleraces}). However, this assembly consists of some thermally and mechanical fragile components, such as the WPA and the wire-grid 
PBS\footnote{These are 0.7~mm thick and have an aperture of $78\times90$~mm aperture.}. So these need to be carefully mounted to minimize stresses due to the mounting procedure as well as during their operation due to temperature and telescope pointing variations. Also, in this subsystem, all the four beams (O1, O2, E1 and E2) get separated by the optics and becomes a very probable location for stray light contamination between the four beams. 

\begin{figure}
\begin{subfigure}{0.59\textwidth}
    \centering
    \includegraphics[scale = 0.275]{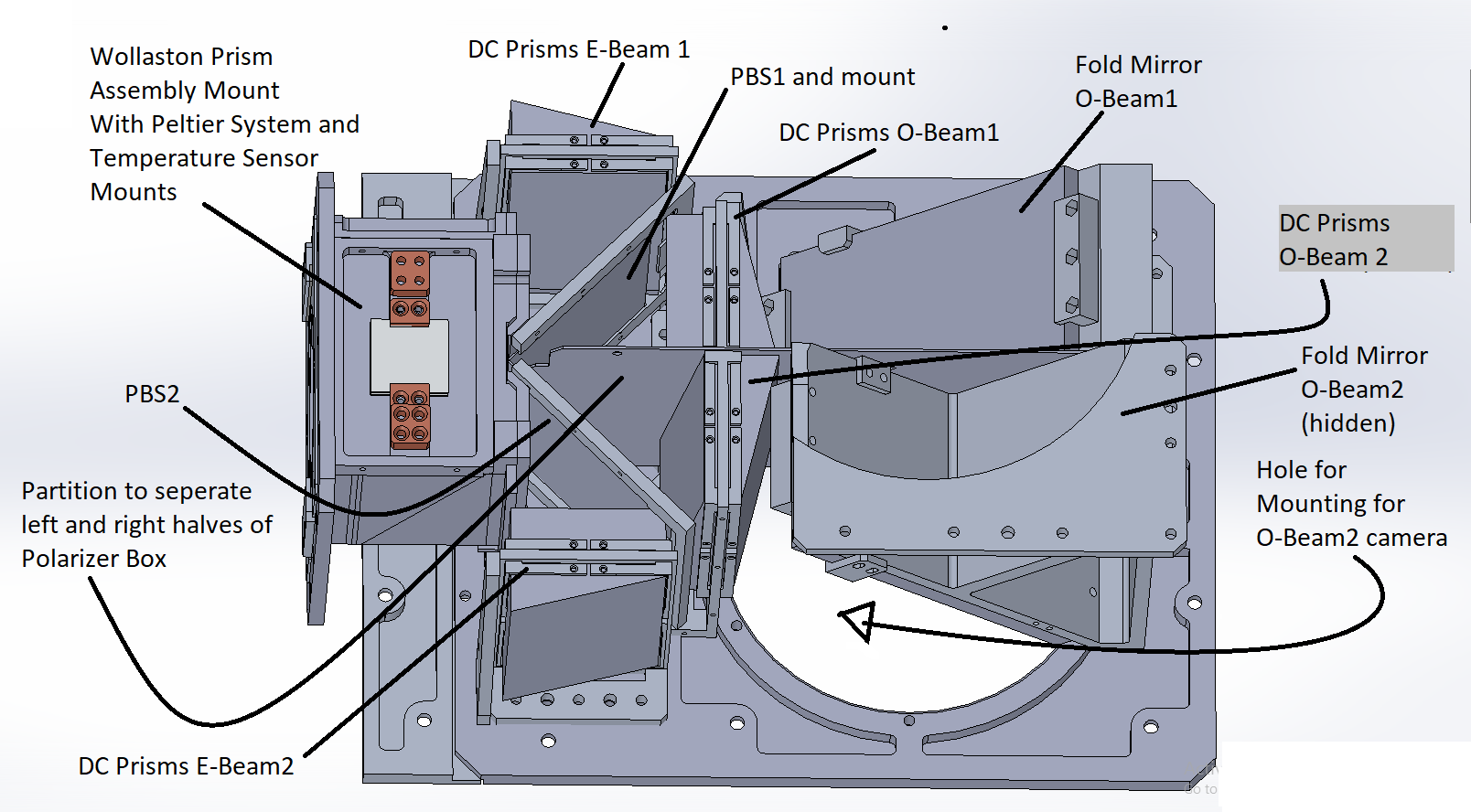}
\end{subfigure}
\begin{subfigure}{0.39\textwidth}
    \centering
    \includegraphics[scale=0.12]{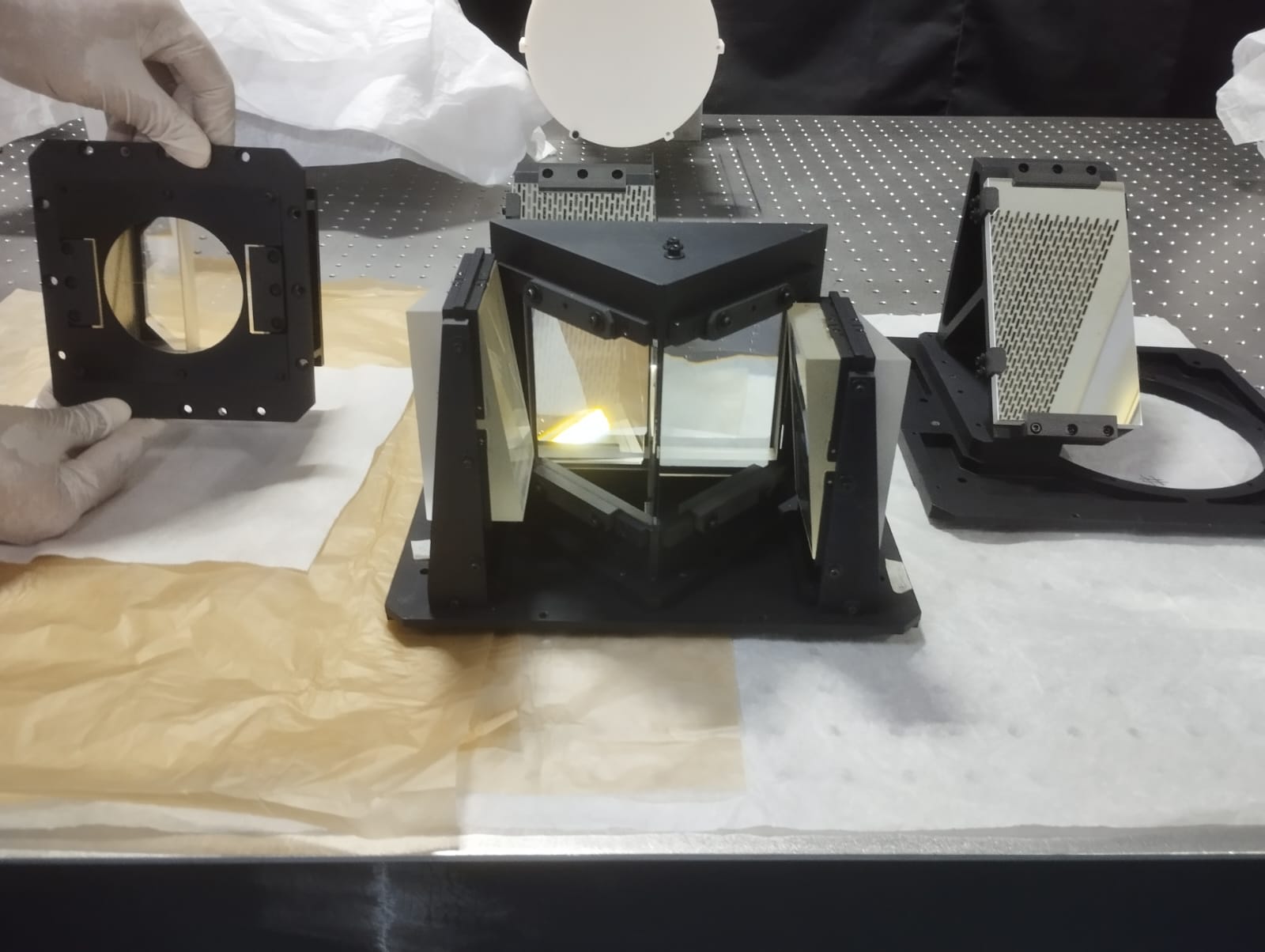}
\end{subfigure}
\caption{Left:Optomechanical Design of the polarizer assembly. Right: Assembled subsystems of the polarizer assembly in the lab.}
\label{polarizer_assembly_optmech}
\end{figure}

Figure~\ref{polarizer_assembly_optmech} shows the overall optomechanical model of the polarizer box. All components, except for one Fold Mirror- the WPA, PBS, Dispersion Corrector Prisms (DC Prisms) and a Fold Mirror are mounted on one plate (base plate). This was done to make alignment and integration of the assembly easier. The other Fold Mirror will be mounted on a plate opposite to the base plate, and its tolerances are not critical. Figure~\ref{polarizer_assembly_optmech} (Right) shows the mounted optics of this assembly before their integration into the Polarizer Box.

The WPA will be inserted in its mount and retained with Teflon strips to contain all the degrees of freedom while also minimizing the mounting stress on the optics. When temperature changes occur, these strips will allow the WPA to expand and contract with minimal stress. Further, the WPA system will be thermally regulated by a Peltier system. Provision for attaching PT100 sensors have been made to measure the real time temperature of the WPA. The PBS is made of Corning Eagle XG glass with a CTE of  $3.2\times10^{-6}/K$\footnote{Data sheet for Corning Eagle XG glass: \newline \href{http://www.matweb.com/search/datasheet.aspx?matguid=772db50a5aca4256b6217449132e859c&ckck=1}{http://www.matweb.com/search/datasheet.aspx?matguid=772db50a5aca4256b6217449132e859c\&ckck=1}}. These are thin and fragile, and will be softly clamped on their mounts using teflon strips. The DC prisms will be mounted on flexure mounts, similar to the mounting system of the lenses. On each rim face, two flexure arms will be used to bond the prisms to their mounts. The two DC prisms for each camera are mounted on either side of a common mounting plate. Lastly, the Fold Mirrors are simply clamped using teflon strips on their mounts, as centering alignment is not a concern for them. To prevent stray light reflected from the mounts as well as minimize leaking of one beam into another beam's camera, a thin metal sheet partition has been placed between the optics of O1 and O2 optics holders (left and right halves of the polarizer assembly box).

\subsection{Design of Truss Structure}
Based on the tolerance analysis of the system, we aimed to limit the overall flexure of the system (from instrument beginning to the polarizer box, which is 1.7~m in length) to within 200 microns when the instrument is pointing to a limiting airmass of 2. This turned out to be a major challenge: designing a truss system which was sufficiently stiff and but also light enough to limit the mass of the instrument to 200 Kgs. 

\begin{figure}
\begin{subfigure}{0.49\textwidth}
    \centering
    \frame{\includegraphics[scale = 0.45]{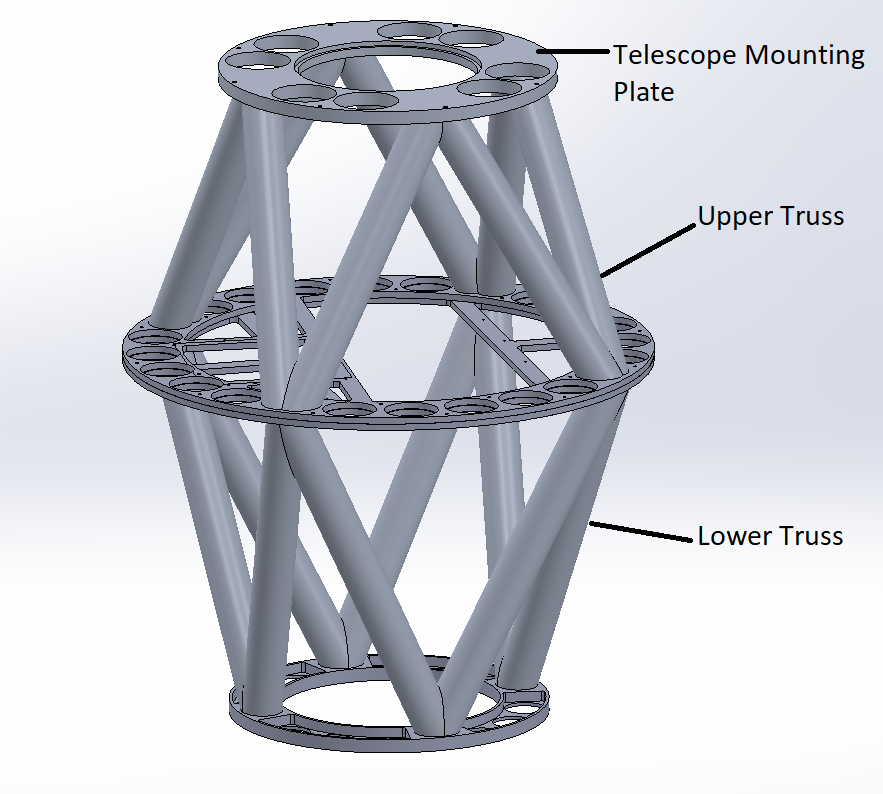}}
    \caption{Truss frame}
    \label{truss_frame}
\end{subfigure}
\begin{subfigure}{0.49\textwidth}
    \centering
    \frame{\includegraphics[scale = 0.75]{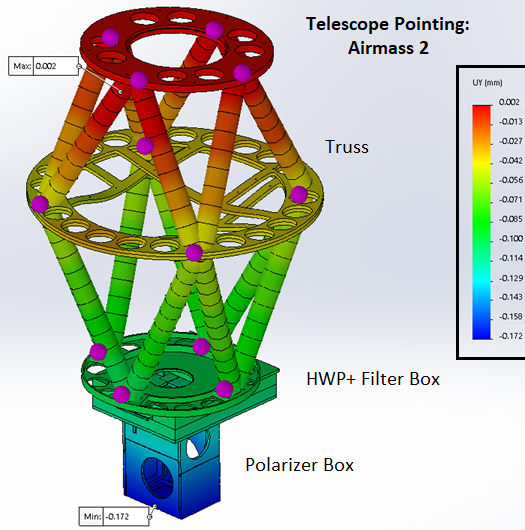}}
    \caption{Simulation of flexure when pointing at airmass of 2.}
    \label{truss_simulation}
\end{subfigure}
\caption{Left: Model of the WALOP-South truss system. Right: Finite Element Analysis (FEA) simulation of instrument flexure with the truss. The instrument is pointing at an airmass of 2. All the sub-components (cameras, collimator etc.) were replaced by suitable forces and moments for the simulation.}
\label{truss_system}
\end{figure}
FEA simulations were done to arrive at the truss design. These simulations were carried out for different pointing airmasses, from the 1 to 2.  Figure~\ref{truss_system} shows the model of the truss system as well a displacement plot in the direction of gravity when the telescope is pointing to airmass of 2. For the analysis, mass of the the individual assemblies (camera, collimator etc.) were replaced by suitable forces and moments. The truss system itself is made of circular Aluminium pipes and consists of two halves- the upper truss connects the mounting flange to roughly the telescope focal plane and the lower truss connects the upper truss to the HWP+Filter mechanism.   

\subsection{Motion Mechanisms and Control Systems}\label{mechanisms}
\begin{figure}
    \centering
\begin{subfigure}{0.46\textwidth}
    \centering
    \frame{\includegraphics[scale = 0.2475]{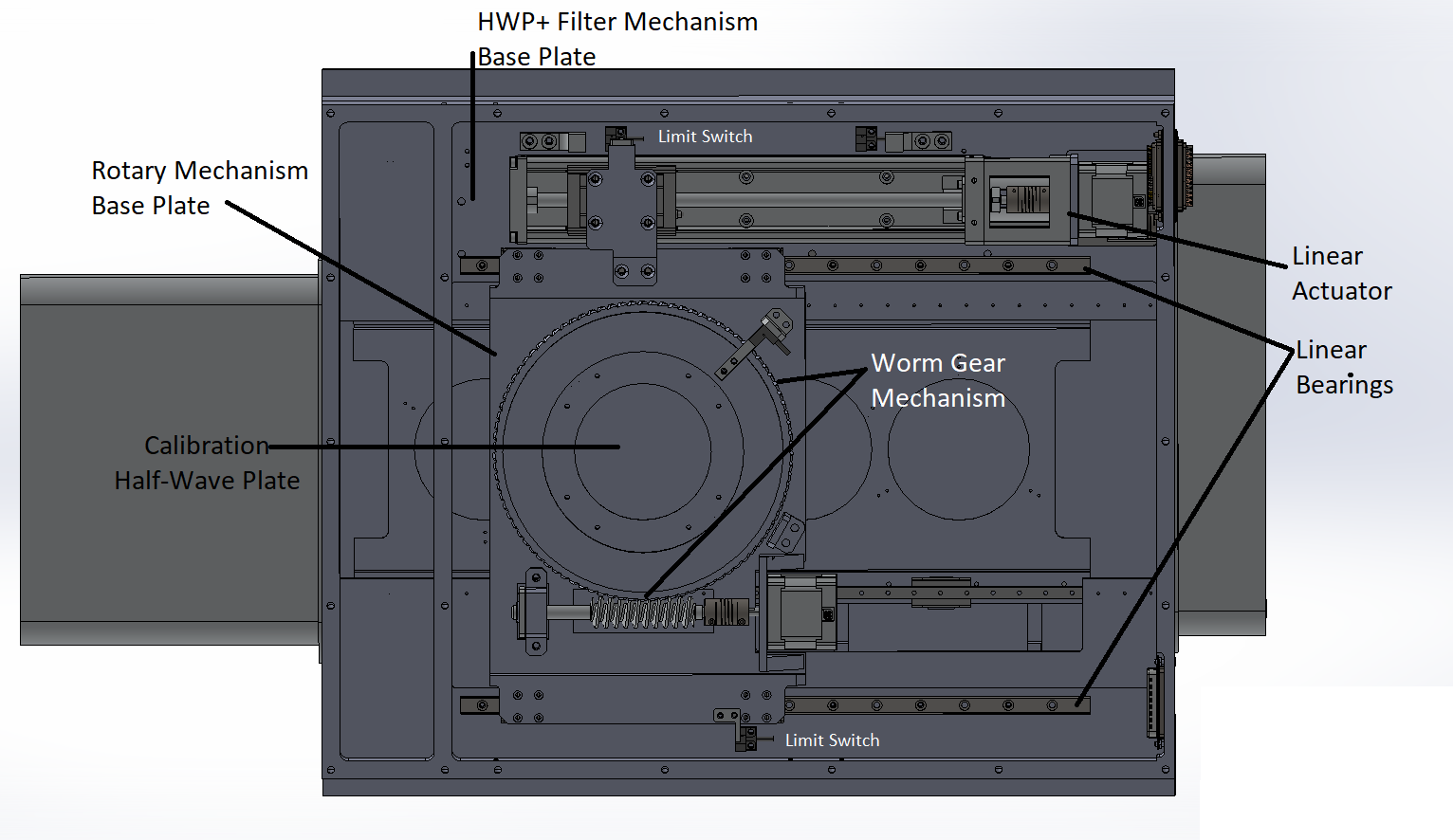}}
\end{subfigure}
\begin{subfigure}{0.52\textwidth}
    \centering
    \frame{\includegraphics[scale = 0.27]{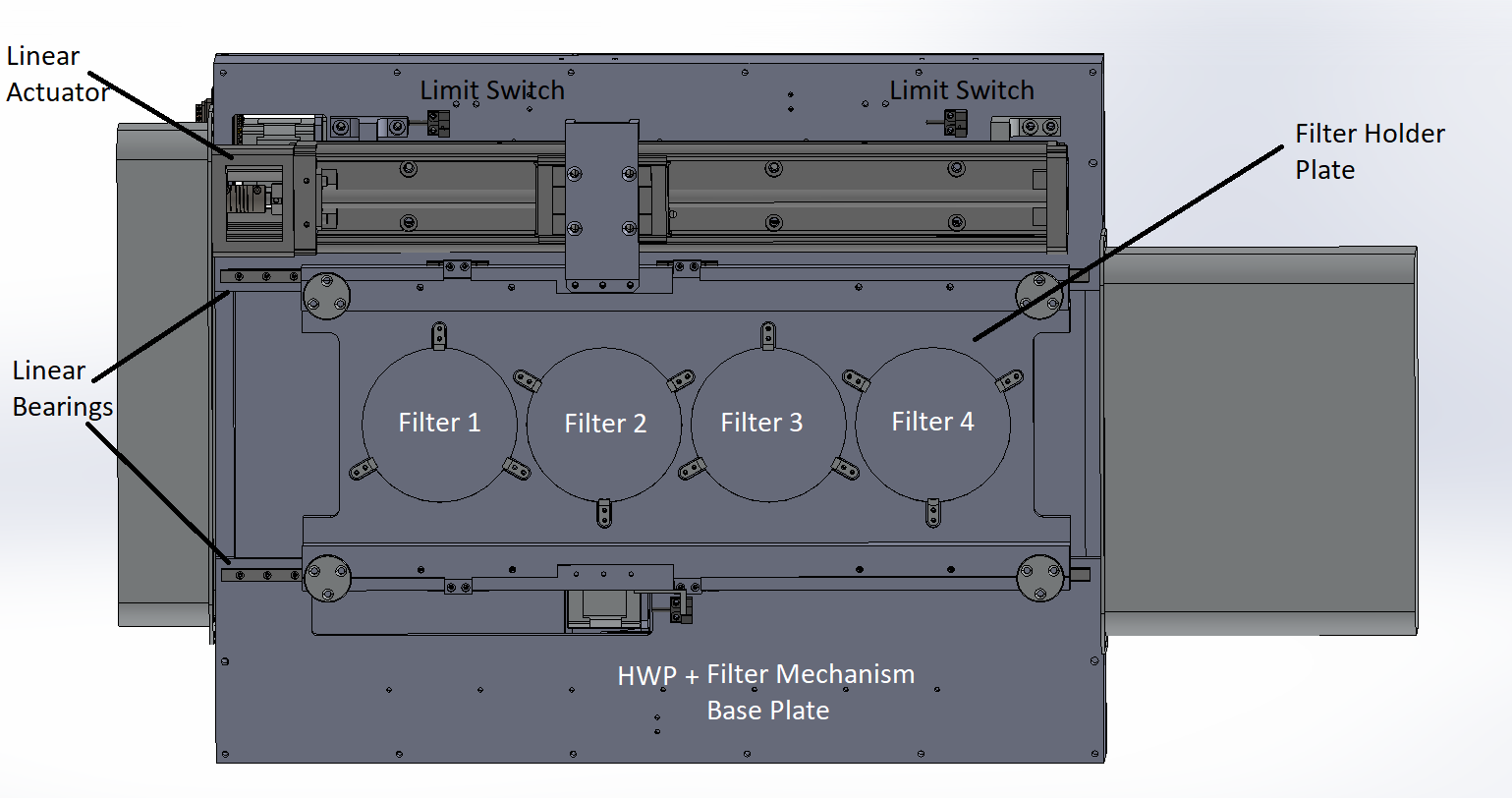}}
\end{subfigure}
\caption{Left: Calibration Half-wave plate assembly optomechanical model layout. Right: Filter Mechanism optomechanical model layout.}
\label{filter_hwp_mechanism}
\end{figure}

\begin{figure}
    \centering

\begin{subfigure}{0.4\textwidth}
    \centering
    \frame{\includegraphics[scale = 0.24]{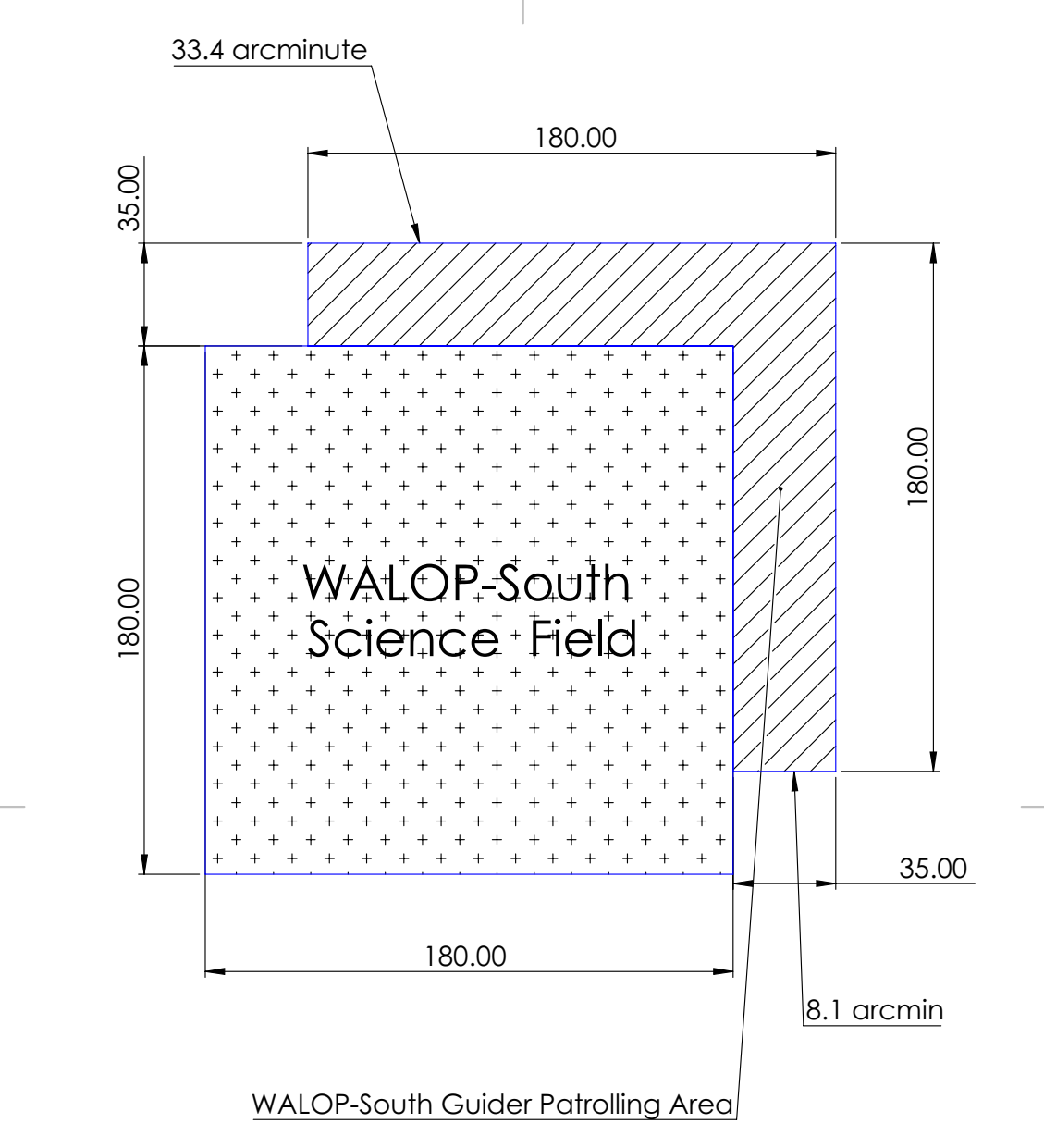}}
\end{subfigure}
\begin{subfigure}{0.59\textwidth}
    \centering
    \frame{\includegraphics[scale = 0.24]{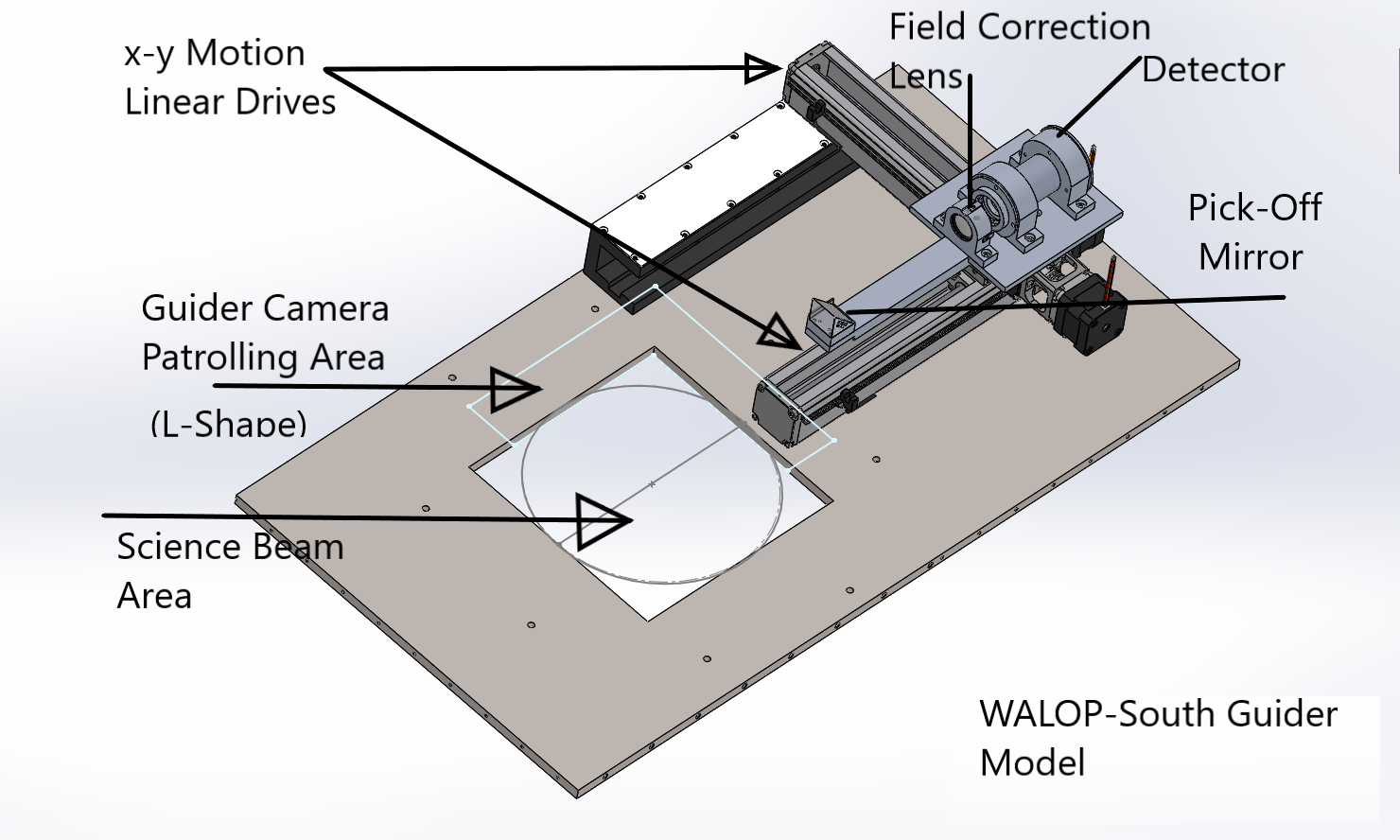}}
\end{subfigure}
    \caption{Left: Patrolling field of guider camera, Right: Optomechanical model of the guider camera.}
    \label{WALOP_S_Guider_optomechanical}
\end{figure}

Table~\ref{control_goals} provides a list of all the control systems included in the WALOP-South instrument. The  calibration linear polarizer (CLP) and the calibration HWP require accurate and controlled rotations when in the optical path with a provision of being shifted away from the optical path when not being utilized. The polarizer sheet is mounted on worm-gear rotation stage, which itself is then mounted on a linear stage. During calibration observations, the CLP will be inserted into the optical path through the linear mechanism. The calibration Half-Wave Plate system, as shown in Figure~\ref{filter_hwp_mechanism}, works in a similar manner. The range and accuracy of the linear and rotary motions of these two systems are listed in Table~\ref{motion_accuracy}. Figure~\ref{filter_hwp_mechanism} also shows the layout of the filter assembly. This assembly has provision for mounting four filters on a plate. The desired filter required for operation with the instrument can be inserted in the optical path through a linear mechanism, similar that used in the CLP and HWP assemblies. All rotary motions are carried out through worm-gear systems as they have zero backlash, allowing better rotational positioning accuracy than spur gears. A common electronic shutter is placed within the collimator barrel for all four cameras as shown in Figure~\ref{collimator_camera_barrel_cross_section}. The guider camera shown in Figure~\ref{WALOP_S_Guider_optomechanical} (please refer to Paper I), has two linear stages on which the entire camera optics is mounted to patrol an L-shaped patrolling area of nearly 500 arcminutes to find a suitable guide star. 

\par The WPA is the most expensive, delicate and temperature sensitive optics in the instrument. We will use a Peltier system to keep the WPA at a constant temperature of $23^{\circ}~C$ throughout its operational lifetime- this is the temperature at which it was bonded with Norland-65 cement. This offers an extra safety feature to protect the WPA from thermal stresses, preventing damage and minimizing stress birefringence that may impact its polarimetric performance.
\par
\begin{figure}
    \centering
\begin{subfigure}{0.29\textwidth}
    \frame{\includegraphics[scale = 0.47]{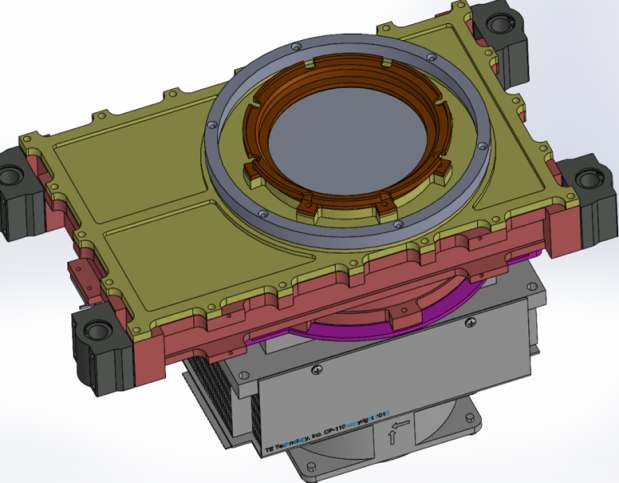}}
\end{subfigure}
\begin{subfigure}{0.365\textwidth}
    \centering
   \frame{\includegraphics[scale = 0.58]{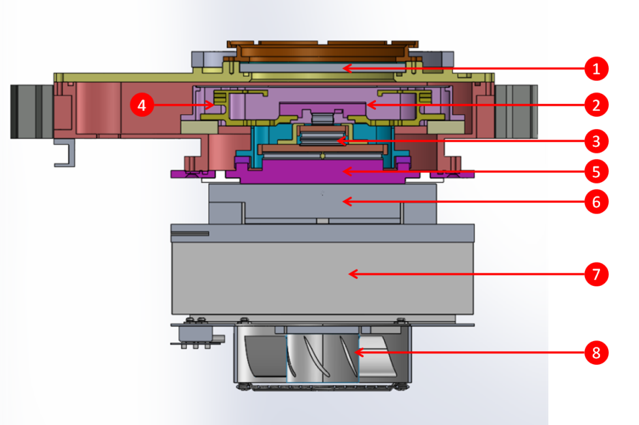}} 
\end{subfigure}
\begin{subfigure}{0.315\textwidth}
    \frame{\includegraphics[scale = 0.445]{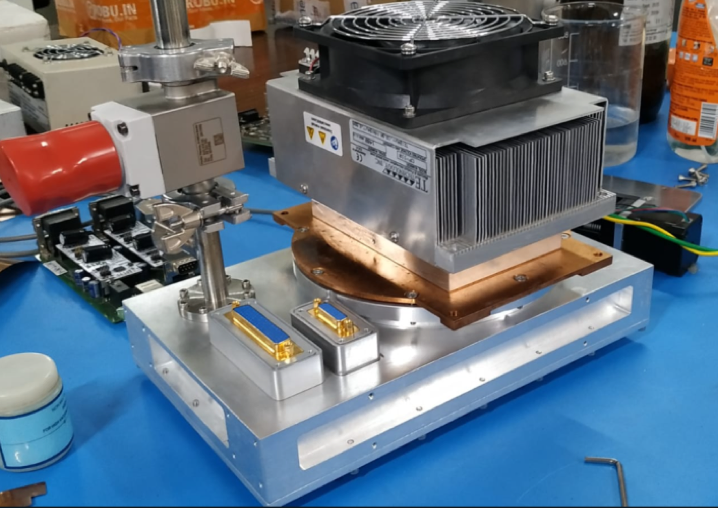}} 
\end{subfigure}

\caption{Left and Middle: CAD model of the CCD dewar. Right: Assembled model in the lab for testing and verification. The parts of the assembly annotated in CAD model are: 1.	Optical window, 2.	CCD , 3. Multistage TEC, 4.	Radiation shield, 5. back-plate, 6.	Cooling plate (outside of the dewar), 7. Heat sink, and 8. fan.}
\label{dewar_model}
\end{figure}

 Each WALOP-South dewar contains a $4k\times4k$ E2V CCD (Figure~\ref{dewar_model}) that is kept at a temperature of $-100~C$ utilizing a thermo-electric cooling system. Dewars for individual cameras are placed on a linear drive system to enable movement along the optical axis for focusing purposes.

\begin{table}[htbp!]
\centering
\caption{Details of the various control systems used in WALOP-South instrument.}
\label{control_goals}
\begin{tabular}{ccccccc}
\hline
Serial No. & Subsystem                                                                   & \begin{tabular}[c]{@{}c@{}}Control\\  Required\\ 1\end{tabular}   & \begin{tabular}[c]{@{}c@{}}Control\\  Required\\ 2\end{tabular} & \begin{tabular}[c]{@{}c@{}}Control \\ Required\\ 3\end{tabular} & \begin{tabular}[c]{@{}c@{}}No. of \\ Motions\end{tabular} & Location                                                               \\ \hline
1          & \begin{tabular}[c]{@{}c@{}}Calibration \\ Polarizer\end{tabular}            & In-out                                                            & Rotation                                                        & -                                                               & 2                                                         & \begin{tabular}[c]{@{}c@{}}Guider + Cal.\\ Polarizer Box\end{tabular}  \\ 
2          & \begin{tabular}[c]{@{}c@{}}Auto-Guider \\ Camera \\ Patrolling\end{tabular} & \begin{tabular}[c]{@{}c@{}}X-direction \\ motion\end{tabular}     & \begin{tabular}[c]{@{}c@{}}Y-direction\\  motion\end{tabular}   & -                                                               & 2                                                         & \begin{tabular}[c]{@{}c@{}}Guider + Cal.\\ Polarizer Box\end{tabular}  \\ 
3          & \begin{tabular}[c]{@{}c@{}}Auto-Guider\\  Camera\end{tabular}               & Filter in-out                                                     & \begin{tabular}[c]{@{}c@{}}Exposure \\ Control\end{tabular}     & -                                                               & 1                                                         & \begin{tabular}[c]{@{}c@{}}Guider + Cal. \\ Polarizer Box\end{tabular} \\ 
4          & \begin{tabular}[c]{@{}c@{}}Half Wave\\  Plate\end{tabular}                  & In-out                                                            & Rotation                                                        & -                                                               & 2                                                         & \begin{tabular}[c]{@{}c@{}}HWP + Filter \\ Wheel Box\end{tabular}      \\ 
5          & Filter Wheel                                                                & Rotation                                                          & -                                                               & -                                                               & 1                                                         & \begin{tabular}[c]{@{}c@{}}HWP + Filter \\ Wheel Box\end{tabular}      \\ 
6          & Shutter                                                                     & Open/Close                                                        & -                                                               & -                                                               & -                                                         & \begin{tabular}[c]{@{}c@{}}Collimator \\ Barrel\end{tabular}           \\ 
7          & \begin{tabular}[c]{@{}c@{}}Wollaston \\ Prism \\ Assembly\end{tabular}      & \begin{tabular}[c]{@{}c@{}}Temperature \\ Control\end{tabular}    & -                                                               & -                                                               & -                                                         & Polarizer Box                                                          \\ 
8          & Dewar 1                                                                     & \begin{tabular}[c]{@{}c@{}}Linear Focus \\ Mechanism\end{tabular} & \begin{tabular}[c]{@{}c@{}}Temperature \\ Control\end{tabular}  & \begin{tabular}[c]{@{}c@{}}CCD \\ Readout\end{tabular}          & 1                                                         & \begin{tabular}[c]{@{}c@{}}Camera 1 \\ Dewar\end{tabular}              \\ 
9          & Dewar 2                                                                     & \begin{tabular}[c]{@{}c@{}}Linear Focus \\ Mechanism\end{tabular} & \begin{tabular}[c]{@{}c@{}}Temperature \\ Control\end{tabular}  & \begin{tabular}[c]{@{}c@{}}CCD \\ Readout\end{tabular}          & 1                                                         & \begin{tabular}[c]{@{}c@{}}Camera 2 \\ Dewar\end{tabular}              \\ 
10         & Dewar 3                                                                     & \begin{tabular}[c]{@{}c@{}}Linear Focus \\ Mechanism\end{tabular} & \begin{tabular}[c]{@{}c@{}}Temperature \\ Control\end{tabular}  & \begin{tabular}[c]{@{}c@{}}CCD \\ Readout\end{tabular}          & 1                                                         & \begin{tabular}[c]{@{}c@{}}Camera 3 \\ Dewar\end{tabular}              \\ 
11         & Dewar 4                                                                     & \begin{tabular}[c]{@{}c@{}}Linear Focus \\ Mechanism\end{tabular} & \begin{tabular}[c]{@{}c@{}}Temperature\\  Control\end{tabular}  & \begin{tabular}[c]{@{}c@{}}CCD \\ Readout\end{tabular}          & 1                                                         & \begin{tabular}[c]{@{}c@{}}Camera 4 \\ Dewar\end{tabular}              \\ \hline
\end{tabular}

\end{table}

\begin{table}[htbp!]
\caption{Required accuracy of the motion systems.}
\label{motion_accuracy}
\centering
\begin{tabular}{cccc}
\hline
Mechanism                        & Type   & Accuracy & Range \\ \hline
Calibration Polarizer Assembly Linear Motion & Linear & 0.1 mm & 300~mm   \\ 
Calibration Polarizer Assembly Rotary Motion & Rotary & 0.01 deg & $360^{\circ}$\\ 
Calibration HWP Assembly Linear Motion       & Linear & 0.1 mm &  120~mm \\ 
Calibration HWP Assembly Rotary Motion       & Rotary & 0.01 deg & $360^{\circ}$\\ 
Dewar Focus Linear Motion       & Linear & 0.005 mm &  10~mm \\ \hline
\end{tabular}

\end{table}

\subsection{CCD Cryostat and Controller Design}
WALOP-South and North will employ thermo-electrically cooled (TEC) CCD cameras. These have been developed in-house at the IUCAA lab. In comparison to liquid-nitrogen cooled CCD cameras, a TEC dewar system is more compact, lightweight, and requires minimal maintenance, making it suitable for WALOP-like instrument designed for automated and long-duration survey. The significant challenges in construction of the TEC-based dewar included selecting a TEC with the appropriate $\delta$T with desired heat pumping capacity, improving the thermal conductivity of the entire system, implementing techniques to reduce parasitic heat loads, and designing a suitable mechanical system to achieve and maintain a vacuum of $10^{-5}$ mBar.
\par The four large CCDs will be read out by the IUCAA Digital Sampler Array Controller (IDSAC) developed in-house by the IUCAA lab\cite{IDSAC}. IDSAC is a versatile CCD Controller with a completely scalable design that provides the capability to manage various detector arrays and mosaics found in modern astronomical instruments. A single IDSAC can control the readout of four CCDs simultaneously. Each CCD is read via four channels at speeds up to 0.5~Megapixels/Sec/Channel at 16-bit resolution. Thus, the CCD readout time per exposure for the WALOP-South's $4k\times4k$ images will be under 20 seconds, which is lower than the typical slew time of the telescope between successive targets, and will minimize the overheads for the PASIPHAE survey.

\section{Assembly and Testing in the Lab}\label{assembly}
\begin{figure}
    \centering
\begin{subfigure}{0.49\textwidth}
    \centering
    \includegraphics[scale=0.185]{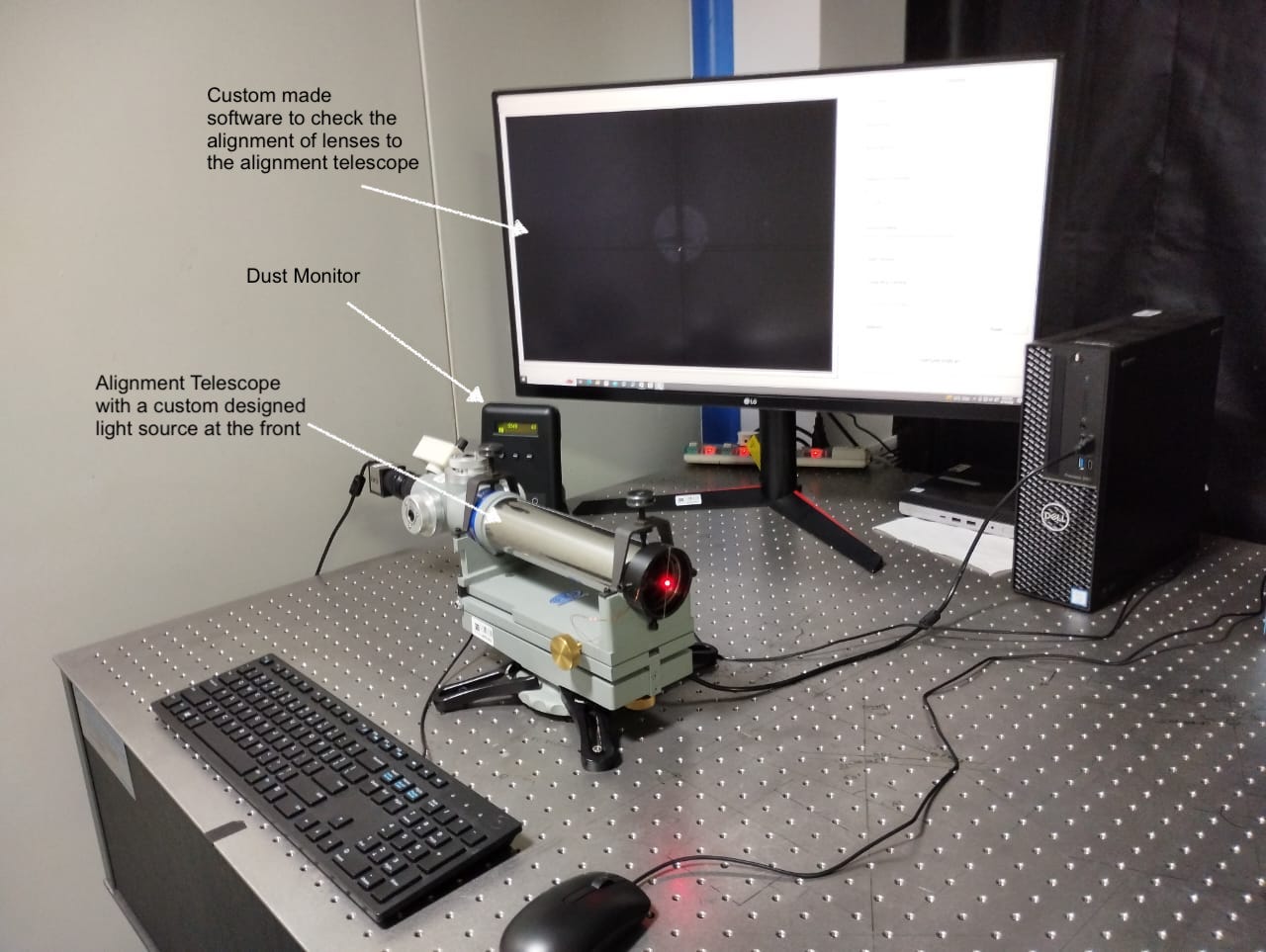}
\end{subfigure}
\begin{subfigure}{0.49\textwidth}
    \centering
    \includegraphics[scale=0.185]{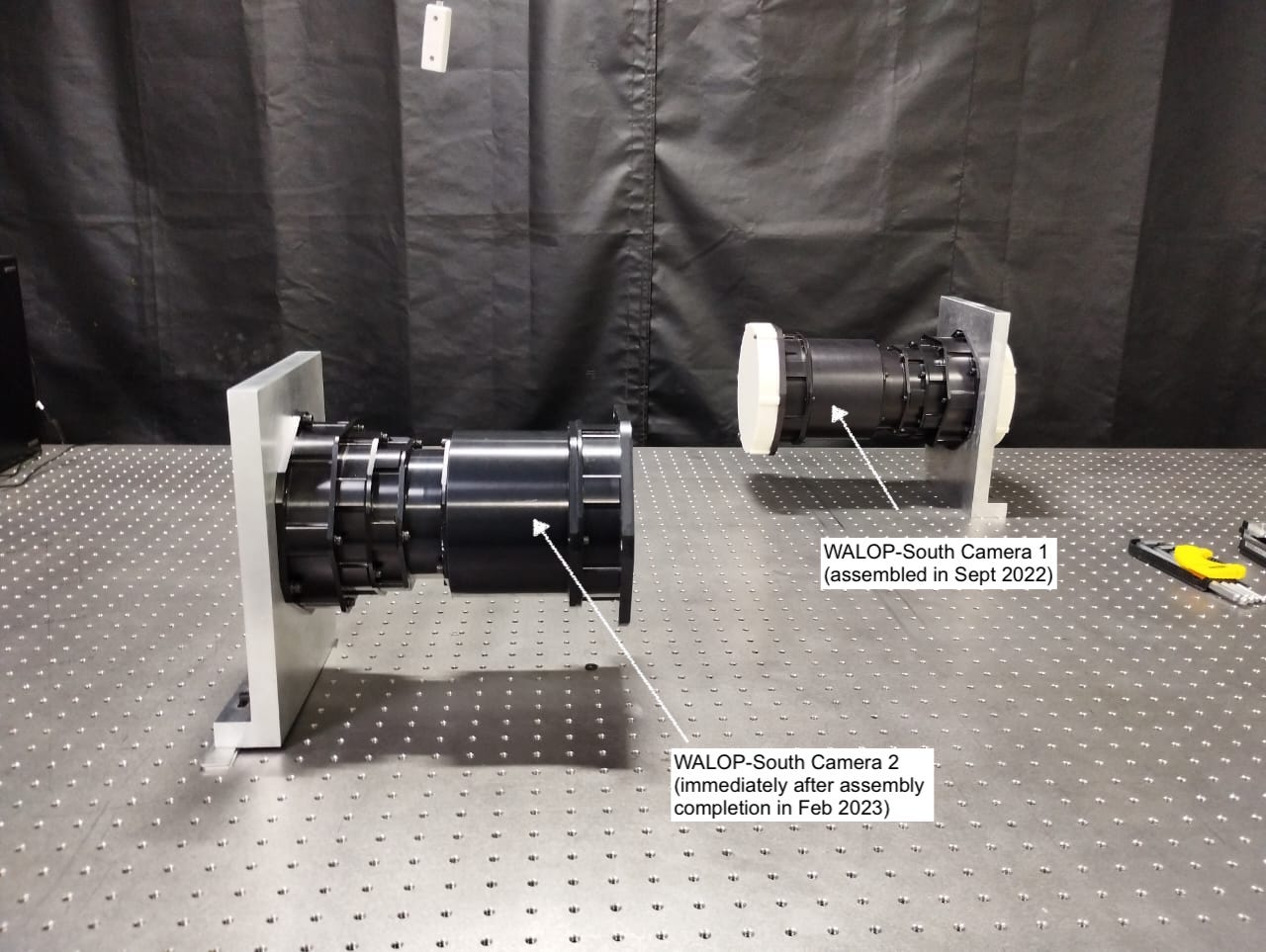} 
\end{subfigure}
\begin{subfigure}{0.49\textwidth}
    \centering
    \includegraphics[scale=0.45]{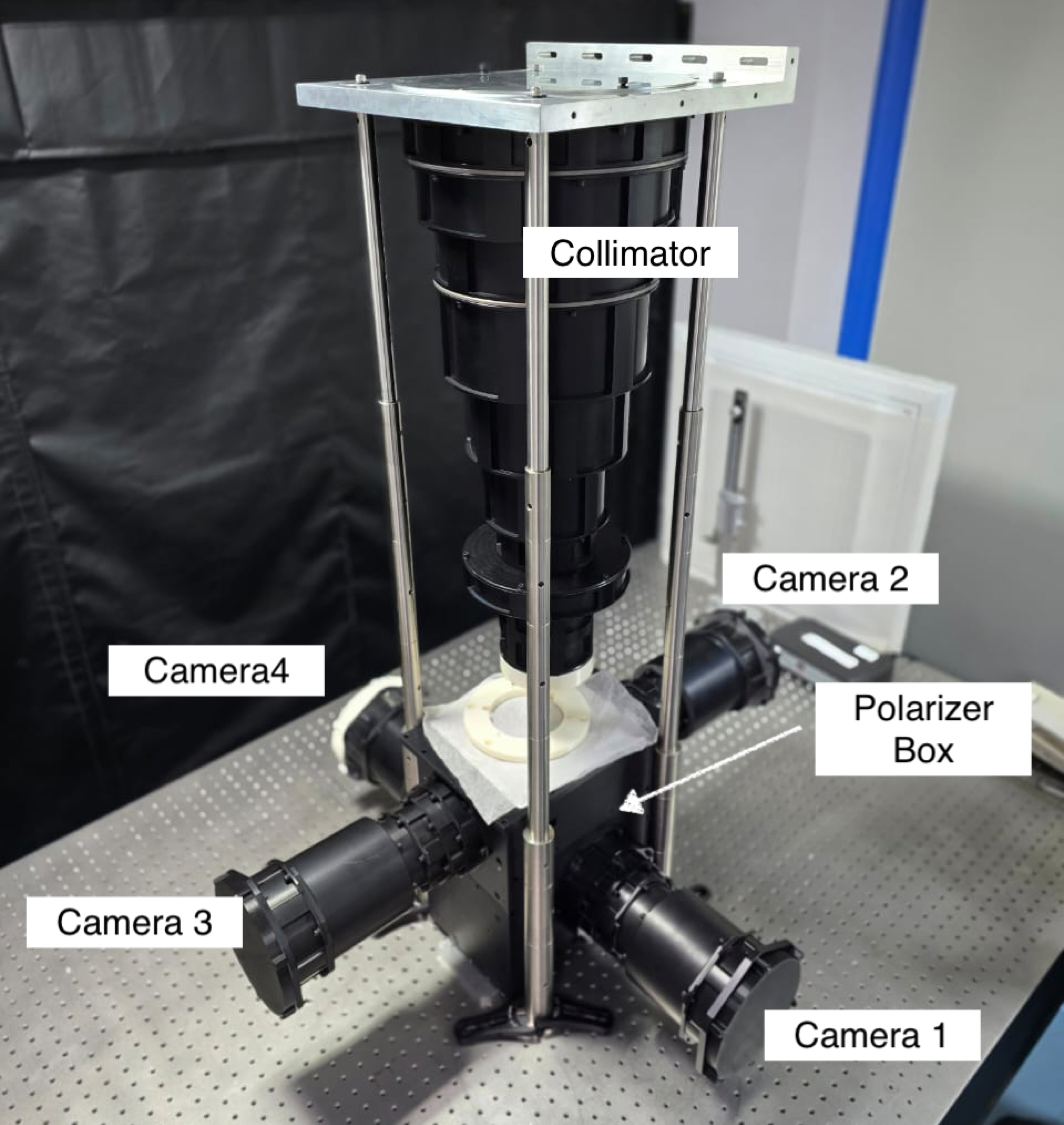}
\end{subfigure}

\caption{Top Left: Alignment telescope set-up used for assembly of the collimator and camera barrels. Top Right: Two of the aligned and assembled cameras. Below: Overall instrument assembly in progress in IUCAA lab; all the major optical systems are mounted.}
\label{real_assembly}
\end{figure}

For accurate assembly and optical alignment of the four cameras and the collimator assembly, an alignment system using an Alignment Telescope (AT) was set up in the lab. It is shown in Figure~\ref{real_assembly}. For the alignment process, a Python-based visual interface was developed that measured the separation between the cross-hair of the AT and the image of the reflected light from the surface of the lenses. As the WALOP-South's optical assemblies are in the form of barrels and the fabricated mounts had a very high precision of $10~\rm{\mu m}$, the assembly process did not pose any significant technical hurdles. It consisted of the following procedure: the first lens of the assembly is mounted on a plate, as shown in Figure~\ref{real_assembly} (top-right). This plate is adjusted so that the tilt and decenter misalignment of the first lens with respect to the AT are corrected by ensuring the image of light from both its surfaces fall on the AT's cross-hair center. Once the first lens is aligned, successive lenses would be mounted and aligned in the above-stated manner. For most lenses, centering adjustments using the mechanical play available between two lens mounts were sufficient for the alignment. For a few lenses, minor tilt corrections were done by inserting shims at appropriate locations.

\par Thus, following this methodology, four cameras and the collimator assembly were assembled. The assembly of the polarizer system, due to more relaxed tolerances, was done relying on the machining accuracies of the components. Figure \ref{real_assembly} (bottom) shows all the optical systems mounted in the lab in a setup.

\par To conduct the preliminary optical performance testing, we assessed the imaging performance of each of the four cameras in conjunction with the collimator assembly. During this test, a grid of square pinholes with a size of 100$\mu m$ was positioned at the nominal focal plane distance from the collimator assembly's first lens to replicate star-like images from the telescope focal plane. Similar to the WALOP-South optical model, this generated a collimated beam that was then directed to any of the four cameras and captured on a detector. By relocating the pinholes across the plane, the full FoV imaging performance testing for the WALOP-South instrument's four cameras was completed. We observed that for all four cameras, and over the entire FoV, the size of the Point Spread Function (PSF) containing 80~\% energy in either the x or y directions was worse than the Zemax predicted value by 10~\% or less. Moreover, the PSF performance remained consistent across the FoV, which is crucial for a wide-field instrument such as WALOP-South.

\par While ideally we need to provide a point source as input to the above collimator+camera optical system for testing the PSF performance, using square pinholes of size 100$\mu m$ is acceptable in our case as the point sources at the focal plane of the telescope are of this size. Due to the slow f/16 beam at the telescope focal plane, the size of a 1.5" object (median seeing FWHM at Sutherland Observatory) is 116.3$\mu m$..

Currently, we are assembling the entire end-to-end optical system of WALOP-South for comprehensive optical and polarimetric testing of the system. Figure~\ref{real_assembly} (bottom) shows such an assembly in progress.

\section{Current Status and Conclusions}\label{conclusions}
In this paper, we have presented an overview of the optical, optomechanical and calibration models for the WALOP instruments, focusing on the WALOP-South instrument.
The technical objectives for the instrument were determined prior to starting the design phase. Several requirements are novel, including: (a) an optical setup featuring four cameras and four-channel polarimetry across a $30\times30$~arcminutes FoV, and (b) accurate polarimetric modeling and calibration to obtain on-sky accuracy of 0.1~\% in $p$. We outline both the challenges and the answers in achieving these design objectives. Thorough analysis was conducted on different systematic factors that could cause measurement errors in the instrument, including stray light, stress birefringence, misalignment of optics due to lens mount tolerances and temperature changes, as well as flexures due to telescope pointing. All of the mentioned impacts were simulated and appropriate plans to lessen their severity were formulated so as to meet the design goals. For instance, a flexible adhesive was selected with care following thermal tests in an environmental testing machine to reduce stress in calcite Wollaston Prisms caused by temperature fluctuations. Another instance is when the lens mounts were created to ensure the alignment of the lens with the instrument is consistent in all temperature and telescope positioning scenarios. The above works have already been published as referred papers\cite{WALOP_South_Optical_Design_Paper, WALOP_Calibration_paper} and previous edition SPIE proceedings\cite{walop_s_spie_2020, walop_stress_birefringence}.


\par The design principles and methods used for creating the WALOP-South instrument can be applied to design other wide-field polarimeters, as shown in the design of the WALOP-North instrument. For instance, larger FoV instruments can be developed by using larger aperture Wollaston Prisms (WP) in a similar optical design, possibly through the use of multiple smaller WPs arranged in a mosaic. Additionally, adjusting the design of WALOPs can allow it to function as a low-resolution imaging spectropolarimeter, utilizing the dispersion caused by calcite WPs. Another instance is the calibration technique created for the WALOP-South device. The model is designed in a way that it is agnostic to the design
detail of the instrument- the model is empirical and can be used for any instrument, making it versatile for use with any instrument. Likewise, the tools and approach created for polarimetric modeling of the device are versatile and can be applied to any polarimeter.


\par Right now, we are putting together the complete optical system of WALOP-South to conduct thorough optical and polarimetric tests on the system in the IUCAA lab. Instrument on-sky commissioning is set to take place in the second half of 2024. We plan to share information about the instrument's performance on the sky, including optical and polarimetric performance, as part of publications after commissioning. Following WALOP-South, the WALOP-North instrument will be commissioned.

\acknowledgments 
The PASIPHAE program is supported by grants from the European Research Council (ERC) under grant agreement No 771282 and No 772253, from the National Science Foundation, under grant number AST-1611547 and the National Research Foundation of South Africa under the National Equipment Programme. This project is also funded by an infrastructure development grant from the Stavros Niarchos Foundation and from the Infosys Foundation. V.P. acknowledges support by the Hellenic Foundation for Research and Innovation (H.F.R.I.) under the “First Call for H.F.R.I. Research Projects to support Faculty members and Researchers and the procurement of high-cost research equipment grant” (Project 1552 CIRCE), funding from a Marie Curie Action of the European Union (grant agreement No. 101107047), and from the Foundation of Research and Technology - Hellas Synergy Grants Program through project MagMASim, jointly implemented by the Institute of Astrophysics and the Institute of Applied and Computational Mathematics. K.T. acknowledges support from the Foundation of Research and Technology – Hellas Synergy Grants Program through project POLAR, jointly implemented by the Institute of Astrophysics and the Institute of Computer Science. This work was supported by NSF grant AST-2109127. I.L., S.K., J. A. K., and N. M. were funded by the European Union ERC-2022-STG - BOOTES - 101076343. Views and opinions expressed are however those of the author(s) only and do not necessarily reflect those of the European Union or the European Research Council Executive Agency. Neither the European Union nor the granting authority can be held responsible for them.
\par S.M. thanks Vinod Vats at Karl Lambrecht Corp. for his inputs and suggestions on various aspects of the Wollaston Prism polarizer assembly design and fabrication, and Prof. Kenneth Nordsieck for sharing his experience and ideas on thermal properties of calcite Wollaston Prisms.
\appendix

\section{Lab Tests on candidate cement for Wollaston Prism Assembly}\label{cement_tests}


\par Calcite WPs are made of two calcite crystal wedges bonded by optical cements such that the optical axis of the two wedges are perpendicular to each other. 
It's coefficients of thermal expansion (CTE) are \cite{Pellicori:70} $ -6 \times 10^{-6}$ and $ +25 \times 10^{-6}$ perpendicular and parallel to the optical axis, respectively. The WALOP polarizer assembly has two calcite Wollaston Prisms (WPs) of aperture $45mm\times80mm$, which needed to be cemented at $23^{\circ}C$.  The temperatures at the Sutherland Observatory and the Skinakas Observatory vary between $40^{\circ}C$ and $-5^{\circ}C$ over the year. The large difference in the CTE values between orthogonal directions leads to large thermal stresses across the interface between the two calcite wedges of a WP. Hence, the bonding material must be very flexible to absorb the thermal stresses over the entire range of temperatures at the observatory. Calcite WPs are known to crack under extreme temperatures if a suitably flexible cement is not used. Although the WALOP WPs will be thermally controlled at around $23^{\circ}C$ during the WALOP operation lifetime, these WPs need to be designed (by using a suitable cement) to survive this temperature range in the event of power/other kinds of failure of the thermal control system for a few hours/a day, till the system can be fixed. 


The only available work done in the past towards finding suitable cements for calcite WP dates back to the 1970s \cite{Pellicori:70}. Using this paper as reference, and after various rounds of discussions with Karl Lambrecht Corporation (KLCC), trials on test glasses by them, as well as discussions with Prof. Ken Nordsieck (PI for the development of polarimetry module on the SALT-RSS instrument that uses calcite WP), we selected the following three optical bonding materials as our candidate cements: (a) Q2-3067, (b) Norland 65 and (c) Norland 73.
Q2-3067 is an optical non-curing gel used for H-POL instrument (University of Wisconsin’s Half-wave Spectropolarimeter) in the early 1990s. Over time, the bonded area acquires some rigidity but never fully solidifies. Norland 65 and 73 are the most flexible cements available with by Norland Products Inc. Both these are curing cements and KLCC was able to use these two cements without any problems. There were other candidate cements that we initially considered, including Dow Corning cements 182, 184 and 93-500. With each of these cements, KLCC faced issues related to bubble removal when contacting surfaces on sample glass pieces. 

To test the suitability of the candidate cements, KLCC fabricated for us 9 small test WPs ($\rm{20mm\times20mm}$ in aperture)- 3 WPs cemented with each of the selected cements. The crystal quality as well as the WP geometry, i.e, the wedge angle in these small WPs is same as that of the WALOP WPs, so the stresses developed will be similar in them. Unless mentioned otherwise, for the rest of this section, the term WP will refer to these test WPs instead of the WALOP WPs. We procured an environmental testing machine (ETM) at the IUCAA lab to carry out the thermal tests on these WPs. Within the ETM chamber, different temperature and humidity environments can be created to test the flexibility of the cements.


3 sets of WPs were created from the total of 9 test WPs obtained from KLCC. Each set comprises of 3 WPs, one WP cemented with each of the candidate cements. Thermal tests were carried out in the ETM on two sets of WPs, called Set 1 and Set 2. The thermal tests done on Set 1 is called Run 1 and that on Set 2 is called Run 2. Run 1 and Run2 are very similar; the main reason for carrying out Run 2 was to test the repeatability of Run 1 results. The details of Run 1 is shown in Table~\ref{run1}. The columns \textit{Chamber Temperature} and \textit{Time Duration} correspond to the chamber temperature (ambient temperature for the WPs) and the time duration for which the WPs were kept in the temperature. After keeping the WPs at each temperature cycle, the chamber was brought back to $25^{\circ}$~C and the WPs were taken out for a thorough visual inspection to identify effects like cracks in the crystals, optical patterns at the interface of wedges due to delamintaion/ breaking of cement bond etc. 

\par The time required for the WPs to reach thermal equilibrium with the ambient temperature (to 1/100 of the temperature difference between the initial WP temperature and the ambient chamber temperature) is around 8 minutes. So a duration of 90-120 minutes inside the chamber is sufficient to observe any damage due to static thermal stresses. Temperature changes are very slow at observatory sites and thermal shocks do not contribute to stresses owing to small timescales required by WPs to reach thermal equilibrium with ambient temperature. After Run1, Set1 of the WPs were kept at $-10^{\circ}C$ for 65 hours to observe term effects of exposure to such low temperatures. The details of Run2 is captured in Table~\ref{run2}. During this run, the chamber was programmed to transition from one temperature state to the next, in 30 minute transition time (without coming to $25^{\circ}$~C). 

\begin{table}[h!]
    \centering
    \begin{tabular}{ccc}
        \hline
        Sr. No. & Chamber Temperature & Time Duration(Minutes) \\
        \hline
        1 & $20^{\circ}C$ & 120 \\
        2 & $15^{\circ}C$ & 120 \\
        3 & $10^{\circ}C$ & 120 \\
        4 & $5^{\circ}C$ & 120 \\
        5 & $-1^{\circ}C$ & 90 \\
        6 & $-6^{\circ}C$ & 90 \\
        7 & $-12^{\circ}C$ & 90 \\
        8 & $40^{\circ}C$ & 120 \\
        \hline
    \end{tabular}
    \caption{Temperature Chamber run details for the first WP set.}
    \label{run1}
\end{table}

\begin{table}[h!]
    \centering
    \begin{tabular}{ccc}
        \hline
        Sr. No. & Chamber Temperature & Time Duration(Minutes) \\
        \hline
        1 & $15^{\circ}C$ & 120 \\
        2 & $5^{\circ}C$ & 120 \\
        3 & $-5^{\circ}C$ & 120 \\
        4 & $-10^{\circ}C$ & 120 \\
        \hline
    \end{tabular}
    \caption{Temperature Chamber run details for the second WP set.}
    \label{run2}
\end{table}

\begin{figure}
    \centering
    \begin{subfigure}[b]{0.49\textwidth}
        \centering
        \includegraphics[scale= 0.51]{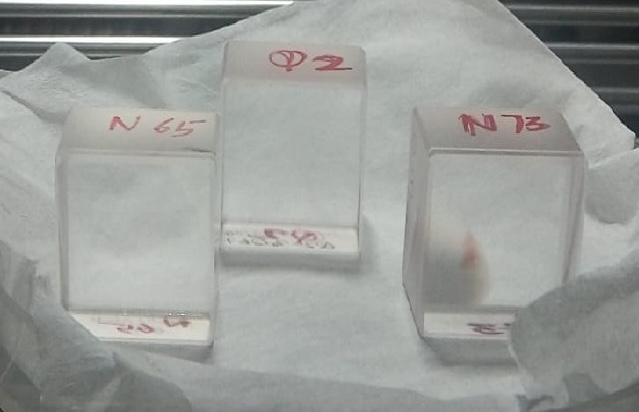}
    \end{subfigure}
    \begin{subfigure}[b]{0.49\textwidth}
        \centering
        \includegraphics[scale= 0.05]{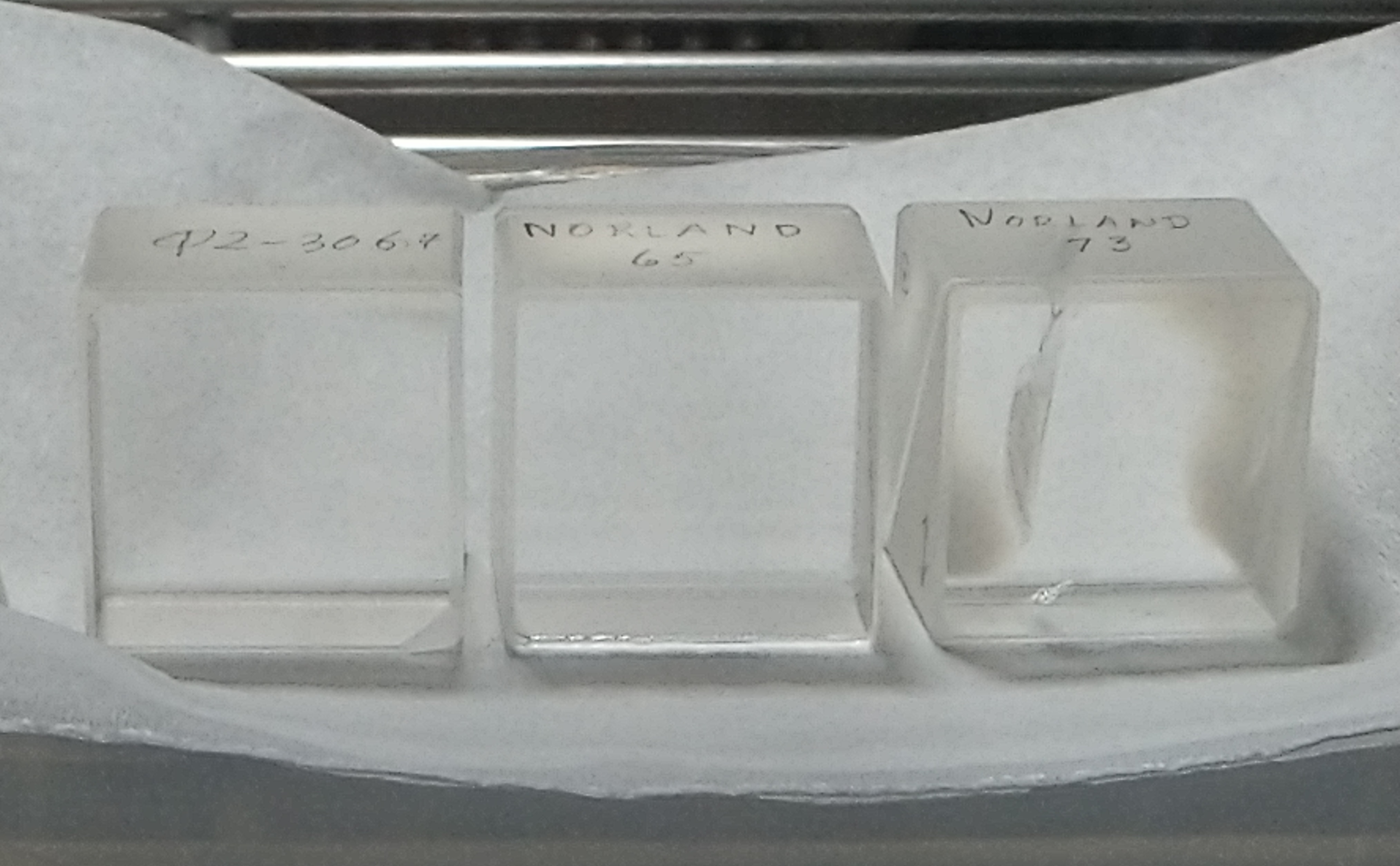}
    \end{subfigure}
        \caption{Left: A translucent layer is formed on WP with Norland 73 cement at lower temperatures ($=< -5^{\circ}C$). Right: Fracture developed in the WP with Norland 73 cement from the second set at $-5^{\circ}C$.}
        \label{N_73_12}
    \end{figure}

Norland 73 cement: During Run1, a translucent layer formed at the cement interface at $-6^{\circ}C$, as shown in Figure~\ref{N_73_12}(Left). This is due to delamination of the cement from thermal stresses. This translucent layer vanishes when the WP is brought back to room temperature, but traces of delamintaion remain, visible in form fingerprint like pattern. During the second run, the WP fractured at $-5^{\circ}C$, as shown in Figure~\ref{N_73_12} (Right).

Norland 65 and Q2-3067 cement: During the first and second run, the WPs remained defect free. For Set1, when exposed to $-10^{\circ}C$ for a duration of 65 hours, for the WP with Norland 65 cement, delamintaion occurred at the edges of the interface. Since this is lower than the minimum temperature expected at the observatories and also because we are designing a thermal control system for the WALOP Polarizer Assembly, the WALOP Wollaston Prisms will not be subjected to such a harsh temperature scenario for more than a few hours.

\textit{Conclusions}: Based on the above tests, Norland 65 was selected for WALOP Polarizer Assembly. Although Q2-3067 is also able to withstand the harsh temperature environments, it is a non-curing cement, and as such it will be very difficult to glue the WALOP Wollaston Prism Assembly consisting of multiple optical elements. 
\bibliography{SPIE2020_Proceedings} 

\begin{thebibliography}{10}

\bibitem{tassis2018pasiphae}
Tassis, K., Ramaprakash, A.~N., Readhead, A. C.~S., Potter, S.~B., Wehus, I.~K., Panopoulou, G.~V., Blinov, D., Eriksen, H.~K., Hensley, B., Karakci, A., Kypriotakis, J.~A., Maharana, S., Ntormousi, E., Pavlidou, V., Pearson, T.~J., and Skalidis, R., ``Pasiphae: A high-galactic-latitude, high-accuracy optopolarimetric survey,'' (2018).

\bibitem{Heiles_2000}
Heiles, C., ``9286 {STARS}: {AN} {AGGLOMERATION} {OF} {STELLAR} {POLARIZATION} {CATALOGS},'' {\em The Astronomical Journal}~{\bf 119},  923--927 (feb 2000).

\bibitem{Panopoulou_2019}
Panopoulou, G.~V., Tassis, K., Skalidis, R., Blinov, D., Liodakis, I., Pavlidou, V., Potter, S.~B., Ramaprakash, A.~N., Readhead, A. C.~S., and Wehus, I.~K., ``Demonstration of magnetic field tomography with starlight polarization toward a diffuse sightline of the ism,'' {\em The Astrophysical Journal}~{\bf 872},  56 (Feb 2019).

\bibitem{Pelgrims2023}
{Pelgrims}, V., {Mandarakas}, N., {Skalidis}, R., {Tassis}, K., {Panopoulou}, G.~V., {Pavlidou}, V., {Blinov}, D., {Kiehlmann}, S., {Clark}, S.~E., {Hensley}, B.~S., {Romanopoulos}, S., {Basyrov}, A., {Eriksen}, H.~K., {Falalaki}, M., {Ghosh}, T., {Gjerl{\o}w}, E., {Kypriotakis}, J.~A., {Maharana}, S., {Papadaki}, A., {Pearson}, T.~J., {Potter}, S.~B., {Ramaprakash}, A.~N., {Readhead}, A.~C.~S., and {Wehus}, I.~K., ``{The first degree-scale starlight-polarization-based tomography map of the magnetized interstellar medium},'' {\em A\&A}~{\bf 684},  A162 (Apr. 2024).

\bibitem{WALOP_South_Optical_Design_Paper}
Maharana, S., Kypriotakis, J.~A., Ramaprakash, A.~N., Rajarshi, C., Anche, R.~M., Shrish, S., Blinov, D., Eriksen, H.~K., Ghosh, T., Gjerløw, E., Mandarakas, N., Panopoulou, G.~V., Pavlidou, V., Pearson, T.~J., Pelgrims, V., Potter, S.~B., Readhead, A. C.~S., Skalidis, R., Tassis, K., and Wehus, I.~K., ``{WALOP-South: a four-camera one-shot imaging polarimeter for PASIPHAE survey. Paper I—optical design},'' {\em Journal of Astronomical Telescopes, Instruments, and Systems}~{\bf 7}(1),  1 -- 24 (2021).

\bibitem{robopol}
Ramaprakash, A.~N., Rajarshi, C.~V., Das, H.~K., Khodade, P., Modi, D., Panopoulou, G., Maharana, S., Blinov, D., Angelakis, E., Casadio, C., Fuhrmann, L., Hovatta, T., Kiehlmann, S., King, O.~G., Kylafis, N., Kougentakis, A., Kus, A., Mahabal, A., Marecki, A., Myserlis, I., Paterakis, G., Paleologou, E., Liodakis, I., Papadakis, I., Papamastorakis, I., Pavlidou, V., Pazderski, E., Pearson, T.~J., Readhead, A. C.~S., Reig, P., Słowikowska, A., Tassis, K., and Zensus, J.~A., ``{RoboPol: a four-channel optical imaging polarimeter},'' {\em Monthly Notices of the Royal Astronomical Society}~{\bf 485},  2355--2366 (02 2019).

\bibitem{HOWPol}
Kawabata, K.~S., Nagae, O., Chiyonobu, S., Tanaka, H., Nakaya, H., Suzuki, M., Kamata, Y., Miyazaki, S., Hiragi, K., Miyamoto, H., Yamanaka, M., Arai, A., Yamashita, T., Uemura, M., Ohsugi, T., Isogai, M., Ishitobi, Y., and Sato, S., ``{Wide-field one-shot optical polarimeter: HOWPol},'' in [{\em Ground-based and Airborne Instrumentation for Astronomy II}{\nolinebreak\hspace{0.1em}]},  McLean, I.~S. and Casali, M.~M., eds.,  {\bf 7014},  1585 -- 1594, International Society for Optics and Photonics, SPIE (2008).

\bibitem{salt_commisioning}
Potter, S.~B., Nordsieck, K., Romero-Colmenero, E., Crawford, S., Vaisanen, P., Éric Depagne, Buckley, D., Koeslag, A., Brink, J., Hetlage, C., Browne, K., Crause, L., Schier, A., and Allington, J., ``{Commissioning the polarimetric modes of the Robert Stobie spectrograph on the Southern African Large Telescope},'' in [{\em Ground-based and Airborne Instrumentation for Astronomy VI}{\nolinebreak\hspace{0.1em}]},  Evans, C.~J., Simard, L., and Takami, H., eds.,  {\bf 9908},  810 -- 817, International Society for Optics and Photonics, SPIE (2016).

\bibitem{Pellicori:70}
Pellicori, S.~F., ``Optical bonding agents for severe environments,'' {\em Appl. Opt.}~{\bf 9},  2581--2582 (Nov 1970).

\bibitem{WALOP_Calibration_paper}
Maharana, S., Anche, R.~M., Ramaprakash, A.~N., Joshi, B., Basyrov, A., Blinov, D., Casadio, C., Deka, K., Eriksen, H.~K., Ghosh, T., Gjerl{\o}w, E., Kypriotakis, J.~A., Kiehlmann, S., Mandarakas, N., Panopoulou, G.~V., Papadaki, K., Pavlidou, V., Pearson, T.~J., Pelgrims, V., Potter, S.~B., Readhead, A. C.~S., Skalidis, R., Svalheim, T.~L., Tassis, K., and Wehus, I.~K., ``{WALOP-South: a four-camera one-shot imaging polarimeter for PASIPHAE survey. Paper II – polarimetric modeling and calibration},'' {\em Journal of Astronomical Telescopes, Instruments, and Systems}~{\bf 8}(3),  038004 (2022).

\bibitem{sky_polarization}
{Maharana, S.}, {Kiehlmann, S.}, {Blinov, D.}, {Pelgrims, V.}, {Pavlidou, V.}, {Tassis, K.}, {Kypriotakis, J. A.}, {Ramaprakash, A. N.}, {Anche, R. M.}, {Basyrov, A.}, {Deka, K.}, {Eriksen, H. K.}, {Ghosh, T.}, {Gjerløw, E.}, {Mandarakas, N.}, {Ntormousi, E.}, {Panopoulou, G. V.}, {Papadaki, A.}, {Pearson, T.}, {Potter, S. B.}, {Readhead, A. C. S.}, {Skalidis, R.}, and {Wehus, I. K.}, ``Bright-moon sky as a wide-field linear polarimetric flat source for calibration,'' {\em A\&A}~{\bf 679},  A68 (2023).

\bibitem{RoboPol_standards}
{Blinov, D.}, {Maharana, S.}, {Bouzelou, F.}, {Casadio, C.}, {Gjerløw, E.}, {Jormanainen, J.}, {Kiehlmann, S.}, {Kypriotakis, J. A.}, {Liodakis, I.}, {Mandarakas, N.}, {Markopoulioti, L.}, {Panopoulou, G. V.}, {Pelgrims, V.}, {Pouliasi, A.}, {Romanopoulos, S.}, {Skalidis, R.}, {Anche, R. M.}, {Angelakis, E.}, {Antoniadis, J.}, {Medhi, B. J.}, {Hovatta, T.}, {Kus, A.}, {Kylafis, N.}, {Mahabal, A.}, {Myserlis, I.}, {Paleologou, E.}, {Papadakis, I.}, {Pavlidou, V.}, {Papamastorakis, I.}, {Pearson, T. J.}, {Potter, S. B.}, {Ramaprakash, A. N.}, {Readhead, A. C. S.}, {Reig, P.}, {Słowikowska, A.}, {Tassis, K.}, and {Zensus, J. A.}, ``The robopol sample of optical polarimetric standards,'' {\em A\&A}~{\bf 677},  A144 (2023).

\bibitem{Tinyanont2018}
Tinyanont, S., Millar-Blanchaer, M.~A., Nilsson, R., Mawet, D., Knutson, H., Kataria, T., Vasisht, G., Henderson, C., Matthews, K., Serabyn, E., Milburn, J.~W., Hale, D., Smith, R., Vissapragada, S., Santos, L.~D., Kekas, J., and Escuti, M.~J., ``\uppercase{WIRC+P}ol: A low-resolution near-infrared spectropolarimeter,'' {\em Publications of the Astronomical Society of the Pacific}~{\bf 131},  025001 (dec 2018).

\bibitem{Vukobratovich_flexure}
Vukobratovich, D. and Richard, R.~M., ``{Flexure Mounts For High-Resolution Optical Elements},'' in [{\em Optomechanical and Electro-Optical Design of Industrial Systems}{\nolinebreak\hspace{0.1em}]},  Bieringer, R.~J. and Harding, K.~G., eds.,  {\bf 0959},  18 -- 36, International Society for Optics and Photonics, SPIE (1988).

\bibitem{Yoder1}
Yoder, P.~R., ``{Optomechanical Systems Design},'' {\em Optical Engineering}~{\bf 20}(2),  155 -- 155 (1981).

\bibitem{walop_stress_birefringence}
{Anche}, R.~M., {Maharana}, S., {Ramaprakash}, A.~N., {Khodade}, P., {Modi}, D., {Rajarshi}, C., {Kypriotakis}, J.~A., {Blinov}, D., {Eriksen}, H.~K., {Ghosh}, T., {Panopoulou}, G.~V., {Pelgrims}, V., {Skalidis}, R., {Pearson}, T.~J., {Gjerl{\o}w}, E., {Mandarakas}, N., {Pavlidou}, V., {Potter}, S.~B., {Readhead}, A. C.~S., {Tassis}, K., {Basyrov}, A., {Papadaki}, K., {Svalheim}, T.~L., and {Wehus}, I.~K., ``{Stress-induced birefringence in the lenses of Wide-Area Linear Optical Polarimeter-South},'' in [{\em Advances in Optical and Mechanical Technologies for Telescopes and Instrumentation}{\nolinebreak\hspace{0.1em}]},  {\em Society of Photo-Optical Instrumentation Engineers (SPIE) Conference Series} {\bf 12188},  121882C (Aug. 2022).

\bibitem{IDSAC}
{Chattopadhyay}, S., {Chordia}, P., {Ramaprakash}, A.~N., {Burse}, M.~P., {Joshi}, B., and {Chillal}, K., ``{IDSAC-IUCAA digital sampler array controller},'' in [{\em High Energy, Optical, and Infrared Detectors for Astronomy VII}{\nolinebreak\hspace{0.1em}]},  {Holland}, A.~D. and {Beletic}, J., eds., {\em Society of Photo-Optical Instrumentation Engineers (SPIE) Conference Series} {\bf 9915},  991524 (July 2016).

\bibitem{walop_s_spie_2020}
Maharana, S., Kypriotakis, J.~A., Ramaprakash, A.~N., Khodade, P., Rajarshi, C., Joshi, B.~S., Chordia, P., Anche, R.~M., Mishra, S., Blinov, D., Eriksen, H.~K., Ghosh, T., Gjerløw, E., Mandarakas, N., Panopoulou, G.~V., Pavlidou, V., Pearson, T.~J., Pelgrims, V., Potter, S.~B., Readhead, A. C.~S., Skalidis, R., Tassis, K., and Wehus, I.~K., ``{WALOP-South: A wide-field one-shot linear optical polarimeter for PASIPHAE survey},'' in [{\em Ground-based and Airborne Instrumentation for Astronomy VIII}{\nolinebreak\hspace{0.1em}]},  Evans, C.~J., Bryant, J.~J., and Motohara, K., eds.,  {\bf 11447},  1135 -- 1146, International Society for Optics and Photonics, SPIE (2020).

\end{thebibliography}
\bibliographystyle{spiebib} 

\end{document}